\documentclass[a4paper,fleqn,usenatbib]{mnras}

\usepackage{graphicx}	% Including figure files
\usepackage{amsmath}	% Advanced maths commands
\usepackage{amssymb}	% Extra maths symbols
\usepackage{url}
\usepackage{txfonts}
\usepackage[T1]{fontenc}
\usepackage{ae,aecompl}

\title[A MUSE view on the dynamics of Crater/Laevens\,I]{Probing the boundary between star clusters and dwarf galaxies: A MUSE view on the dynamics of Crater/Laevens\,I}
\author[K. Voggel et al.]{Karina Voggel$^{1}$\thanks{E-mail: kvoggel@eso.org}, Michael Hilker$^{1}$, Holger Baumgardt$^{2}$, Michelle L.\,M.\,Collins$^{3}$,\newauthor
Eva K.\,Grebel$^{4}$, Bernd Husemann$^{1}$, Tom Richtler$^{5}$, Matthias J.\,Frank$^{6}$
\\
$^{1}$ European Southern Observatory, Karl-Schwarzschild-Str.~2, 85748 Garching bei M\"unchen, Germany \\
$^{2}$ School of Mathematics and Physics, The University of Queensland, Brisbane, QLD 4072, Australia \\
$^{3}$ Department of Physics, University of Surrey, Guildford, GU2 7XH, Surrey, UK \\
$^{4}$ Astronomisches Rechen-Institut, Zentrum f\"ur Astronomie der Universit\"at Heidelberg, M\"onchhofstr. 12-14, 69120 Heidelberg, Germany \\
$^{5}$ Departamento de Astronom\'{\i}a, Universidad de Concepci\'on, Concepci\'on, Chile \\
$^{5}$ Landessternwarte, Zentrum f\"ur Astronomie der Universit\"at Heidelberg, K\"onigsstuhl 12, 69117 Heidelberg, Germany \\
}

\date{Accepted 2016 April 22; Received 2016 April 15; in original form 2016 March 24.}

\pubyear{2016}

\begin{document}
\label{firstpage}
\pagerange{\pageref{firstpage}--\pageref{lastpage}}
\maketitle

\begin{abstract}
We present MUSE observations of the debated ultra faint stellar system Crater. We spectroscopically confirm 26 member stars of this system via radial velocity measurements. We derive the systematic instrumental velocity uncertainty of MUSE spectra to be 2.27$\rm \,km\,s^{-1}$. This new dataset increases the confirmed member stars of Crater by a factor of 3. One out of three bright blue stars and a fainter blue star just above the main-sequence-turn-off are also found to be likely members of the system. The observations reveal that Crater has a systemic radial velocity of  $v_{\rm sys}=148.18^{\rm +1.08}_{\rm -1.15}\rm \,km\,s^{-1}$, whereas the most likely velocity dispersion of this system is $\sigma_{\rm v}=2.04^{\rm +2.19}_{\rm -1.06}  \rm \,km\,s^{-1}$. The total dynamical mass of the system, assuming dynamical equilibrium is then  $M_{\rm tot}=1.50^{+4.9}_{-1.2}\cdot 10^{\rm 5}M_{\odot}$ implying a mass-to-light ratio of M/L$_{\rm V}$=8.52$^{+28.0}_{-6.5}\, M_{\odot}/L_{\odot}$, which is consistent with a purely baryonic stellar population within its errors and no significant evidence for the presence dark matter was found. We also find evidence for a velocity gradient in the radial velocity distribution. We conclude that our findings strongly support that Crater is a faint intermediate-age outer halo globular cluster and not a dwarf galaxy. 
\end{abstract}


%%%%%%%%%%%%%%%%%%%%%%%%%%%%%%%%%%%%%%%%%%%%%%%%%%

%%%%%%%%%%%%%%%%% BODY OF PAPER %%%%%%%%%%%%%%%%%%

\section{Introduction}
The recent discovery of ultra-faint dwarf spheroidals and extended star clusters has changed our view on small stellar systems. A decade ago, globular clusters and dwarf galaxies were well separated in size-luminosity parameter space  (\citealt{Gilmore2007}), and thus could easily be distinguished from each other.
A multitude of new objects has been gradually filling the gap at the faint end of the radius-magnitude scaling relation, between dwarf spheroidal galaxies and star clusters \citep[e.g.][]{Misgeld2011, Mcconn2012, Willman2005,Huxor2005, Zucker2006, Zucker2006b, Belokurov2007, Belokurov2008, Laevens2015, Martin2015a}. 

With magnitudes of $M_{\rm V}>-$6 and sizes between 10-100\,pc  their structural parameters alone do not allow us to infer their nature. To distinguish the nature of ultra-faint objects one has to study their chemical and dynamical properties in greater detail. Some attempts have been made to clarify what a galaxy is apart from its size, e.g., \citet{Willman2012}, conclude that star formation that lasted for hundreds of Myr and the presence of dark matter are the main discriminators between star clusters and dwarf galaxies in the boundary region. 

For several newly discovered objects in the boundary region spectroscopic follow-up observations were carried out to explore their nature with kinematical data of their stellar populations  \citep[e.g.][]{Martin2015b, Martin2015c, Kirby2015}. 

The dwarf galaxies with the lowest baryon content are expected to be extremely dark matter dominated and to show highly elevated mass-to-light ratios (M/L$_{\rm V}$), whereas globular clusters seem to be dark-matter-free. While the small sizes of globular clusters (a few pc compared to the typically much more extended dwarf spheroidal galaxies) and the impact of the tidal field of the host galaxy  make the detection of dynamical signatures of dark matter harder, detailed velocity dispersion profiles of remote outer halo globular clusters, which are less affected by Galactic tides, suggest that mass follows light \citep[e.g.][]{Jordi2009, Frank2012}. 

%%%%
Crater/Laevens\,I\footnote{For continuity with other recently published work, we will use Crater as naming convention for this paper} is an ultra faint object, independently discovered by \citet{Belokurov2014} and \citet{Laevens2014} in the outermost halo of the Milky Way. Before the discovery by professional astronomers Crater was already identified by the amateur astronomer Pascal Le D\^{u} in the January 2014 issue of the magazine L'Astronomie \footnote{\url{http://www.cielocean.fr/uploads/images/FichiersPDF/L-Astronomie-_Janvier2014.pdf}}.
\citet{Belokurov2014} discovered Crater in observations from the ESO VST ATLAS survey. According to their studies, Craters half-light radius is $r_{\rm h}=30$\,pc and the absolute magnitude is $M_{\rm v}=-5.5$, placing it right at the boundary between extended star clusters and the faintest dwarf galaxies in terms of size and magnitude. Crater's heliocentric distance of 170\,kpc locates it farther than any other previously known Milky Way globular cluster, but well among dwarf galaxy distances. 

The groundbased colour-magnitude diagram of Crater shows that the majority of the stellar population is old (between 7 and 10\,Gyr) and metal-poor, except for a handful of possible luminous ``blue loop" stars. These blue loop stars could be as young as 400\,Myr, and if confirmed as members, would indicate a recent episode of star formation. As an extended star formation history is a diagnostic of dwarf galaxies (e.g. \citealt{Willman2012}), \citealt{Belokurov2014} concluded this newly discovered object must be a dwarf galaxy.

Simultaneously, this faint object was also discovered in the Pan-STARRS1 survey by \citet{Laevens2014}. They measured a slightly fainter absolute magnitude of $M_{\rm V}=-4.3\pm 0.2$ and a slightly smaller heliocentric distance of 145$\pm$17\,kpc, which also results in a smaller half-light radius of $r_{\rm h}=20\pm 2\,$pc. They conclude that Crater consists of a stellar population that is 8-10\,Gyr old and metal-poor with $-2.3\rm \, dex<\left[\rm Fe/H\right]<-1.5$\, dex. In their work the tentative blue loop stars were detected as well, but they argue that these are not part of the system. Combined with their slightly smaller structural parameters they conclude that this object has the typical properties of a young outer halo globular cluster, and thus classify it as such naming it Laevens\,I. 

\cite{Kirby2015} presented spectra of 14 potential member stars of Crater, which were taken with DEIMOS on Keck\,II. They find a heliocentric velocity of $v_{\rm sys}=149.3 \pm \rm1.2\,km\,s^{-1}$ for the 10 sample stars that they consider to be member stars. They derive a velocity dispersion of $\sigma_{\rm v}<4.8\rm \,km\,s^{-1}$  at a confidence level of 95\%. Considering the membership status of the tentative blue loop stars, they find that two of them are not members and a third one is an ambiguous case. This blue star is within their $2.58\, \sigma$ radial velocity membership criterion, but it is excluded as a member based on its position in the CMD. In this work it was also concluded that this object is most likely an outer halo GC.

Another attempt to clarify the nature of Crater was performed by \citet{Bonifacio2015} with X-SHOOTER spectra of two red giant stars in the system. They derive a radial velocity of $V_{1}=144.3\pm 4.0\, \rm km\,s^{-1}$ for the first and $V_{2}=134.1\pm 4.0\rm \,km\,s^{-1}$ for the second probed star. They conclude that both stars are probable members of the stellar system, and that their velocity difference implies a dispersion of $\sigma_{\rm v}>3.7\rm \,km\,s^{-1}$ at 95\% confidence level, if one ignores the errors on the stellar velocities. If these errors are taken into account, their measurement is, however, consistent with a velocity dispersion of 0. They determine metallicities of $[\rm Fe/H]=-1.73$\, dex and $[\rm Fe/H]=-1.67$\, dex for the two stars. Their spectral and photometric data are consistent with an age of 7\,Gyr for the majority of Crater's stellar population, and the blue stars can be interpreted as a population with the same metallicity but an age of only 2.2\,Gyr. Thus, in this work it was concluded that Crater is more likely to be a dwarf galaxy. 

Recently, a new deep HST photometric study of the CMD of Crater (\citealt{Weisz2015}) revealed that its stellar population is well described by a single age of 7.5\,Gyr with a metallicity of [M/H]$\sim$ -1.65\,dex. Similar to \citet{Bonifacio2015} they also detect the blue stars just above the main sequence turnoff. They conclude that the most likely explanation for this sparse population are blue stragglers and not an intermediate age second generation of stars. This result would imply that Crater is a globular cluster, although the majority of intermediate-age star clusters are more metal-rich than Crater. 

One outlier to this trend of shigher metallicities with ages is Lindsay 38 in the SMC, an intermediate age star cluster that is relatively similar to Crater in size and metallicity. It has a measured metallicity of -1.6\,dex and an age of 6.5\,Gyr \citep{Glatt2008}. Lindsay 38 is also comparable in structural parameters to Crater. The half-light radius measured by \citealt{Weisz2015} of Crater is $r_{\rm h}=19.4$\,pc while Lindsay 38 has a half-light radius of  $r_{\rm h}=20.93$\,pc \citep{Glatt2009}.  

The ongoing controversy on the nature of this object, even after spectroscopic follow-up, shows how unclear and blurry the distinction between dwarf galaxies and globular clusters is when we get to the boundary regions where they are not distinguishable any longer by their structural parameters. The location of Crater in the outer Milky Way halo means that the presence of dark matter in this object can be tested, as the tidal field of the galaxy has a much reduced influence at such distances, minimising the tidal effects on the stellar velocities (\citealt{Baumgardt2005}). With seven times the physical size of a typical GC and at the largest galactocentric distance of any GC in the MW halo, it is a unique system to study the formation and evolution mechanisms of ultra-faint dwarfs or intermediate-age, extended globular clusters at low metallicities.

In this paper, we present new radial velocity measurements of stars in Crater using the new IFU instrument MUSE on the VLT.
We perform a detailed study of Craters stellar dynamics by increasing the sample size of likely member stars to 26.
Simultaneously, the IFU data will allow us to characterise the chemical composition of the system, by analysing the abundances of 
absorption features in their spectra. This analysis of the chemistry of Crater will be published in a second work on this object. Throughout the paper we adopt a distance to Crater of $d=145$\,kpc, a magnitude of $M_{\rm V}$=-5.3 and a half-light radius of $r_{\rm h}$=19.4\,pc, unless otherwise noted. All these values are taken from the HST study of \citet{Weisz2015}.

\section{MUSE Observations}
Crater was observed with MUSE (\citealt{Bacon2010}) mounted on UT4 of the VLT, during the night of March 13th 2015, under ESO programme Nr. 094.D-0880 (PI: Hilker). MUSE is an integral field spectrograph with a spatial sampling of 0.2\,$\arcsec$ and a spectral resolution between R=1500-3000 along its wavelength coverage. We used the nominal mode of MUSE which covers the wavelength range of $4750-9300\,\AA$. 	
Three (out of four planned) pointings centered around Crater (R.A.[J2000] = 11:36:17.965, $\delta$[J2000] = -10:52:04.80), with 2$\times$1400\,s each, were observed with the $1\,\rm arcmin^{2}$ field-of-view of MUSE. Each FOV covers Crater out to a spatial extent of roughly one half-light radius. Thus, the three observed pointings cover three quarters of the area within the half-light radius of Crater. In each observation the position angle of the spectrograph was rotated by 90$^{\circ}$ after the first 1400\,s exposure. A dedicated sky field of 100s was taken after each observing block. 

\subsection{Data reduction process}
The MUSE raw data were processed using version 1.0 of the provided ESO pipeline (\citealt{Weilbacher2012}). A master-bias, master-flatfield and a wavelength calibration were generated using the bias, flat and arc-lamp calibrations taken in the same night. The wavelength dependent line spread function was obtained from the calibration files.

Each exposure was bias subtracted and divided by the flatfield. The generated trace tables and the static geometric calibration, as well as the wavelength calibration, were then applied to the data cube. This calibrates the spectra to physical units of wavelength. The results of each basic reduction were then stored in separate \textit{pixtables}, an intermediate product of the pipeline. These were then processed with the muse\_scipost recipe. This recipe applies the on-sky calibrations to the \textit{pixtables}. The astrometric calibrations of the two dimensional positions were applied, the sky was removed and the flux calibration was carried out in this step. This recipe stores one fully reduced \textit{pixtable} for each of our 6 observations. 
\begin{figure}
	\includegraphics[width=\columnwidth]{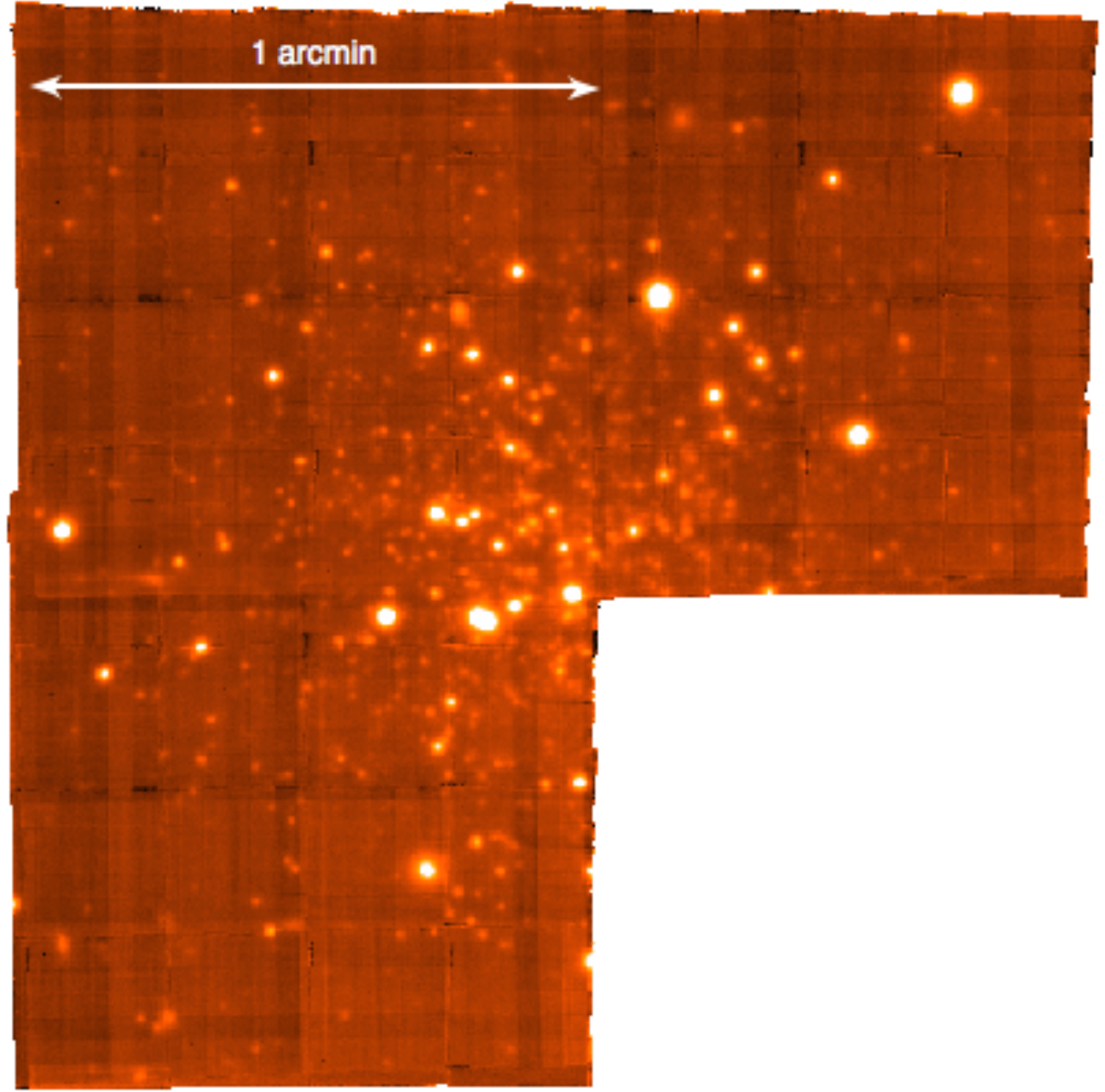}
    \caption{White light image of the final Crater MUSE cube for which all 6 observations have been combined. The cube is collapsed along its spectral axis. The image is oriented with north oriented towards the top and east towards the left.}
    \label{fig:cube}
\end{figure}

The 6 reduced pixtables were then merged into the final datacube using the MUSE exp\_combine recipe of the pipeline. For this recipe the offsets of each exposure were provided manually, as the pipeline is not able to correct slight positional offsets automatically when the observations involve a position angle rotation of 90$^{\circ}$. The position cross-correlations between exposures were provided by fitting the 5 brightest point sources in the FOV of each exposure with a Gaussian function. We then used the weighted mean of the position offsets as the geometric shifts provided to the pipeline. This recipe delivered a final three dimensional \mbox{datacube} with the science data in the first extension of the fits file, and an estimate of the noise in the second extension.
The final MUSE cube, collapsed along its spectral axis, is shown in Fig. \ref{fig:cube}. 

\subsection{Absolute velocity calibration}
\label{sec:abs}
The wavelength calibration for MUSE cubes is performed directly by the pipeline using the provided arc lamp spectra. The pipeline corrects for systematic wavelength shifts between the arclamp lines and the observed sky lines by using the strong sky lines at 5577\,\AA\,  and 6300\,\AA . This correction is propagated linearly to the rest of the spectrum, as to get rid of the wavelength dependency of this correction. It is unknown how well this linear propagation of the corrections holds in the Ca triplet wavelength range.

As we use the near infrared Ca II triplet to measure the radial velocities, and the velocity calibration of the pipeline was done at another wavelength range, we perform our own calibration to get rid of any leftover wavelength dependencies of the correction. For this precise velocity calibration we used a set of bright OH sky lines, located at the same wavelength range as the Ca\,II triplet. 

\begin{figure}
	\includegraphics[width=\columnwidth]{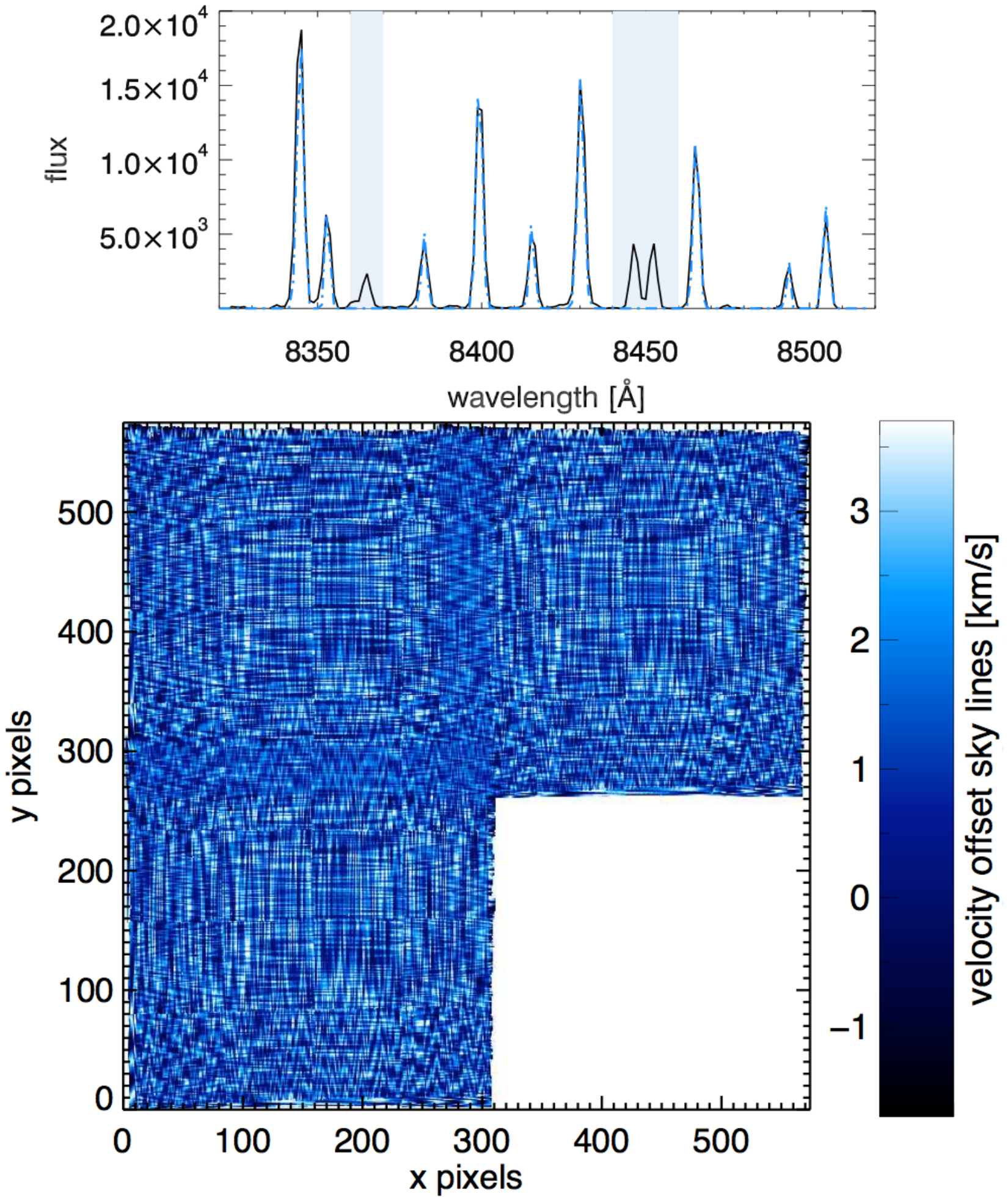}
    \caption{Top panel: The spectrum of the sky emission lines in the Ca triplet wavelength region for the co-added spaxels of the position of star 2. The shifted sky line model is shown in blue. The two shaded regions are excluded from the fit. Bottom panel: The observed MUSE FOV with a colour coding for the velocity corrections. The velocity offset is inferred from the sky lines and determined for each spaxel separately.}
    \label{fig:sky}
\end{figure}

For that we generated a MUSE cube similar to the original reduction except for not subtracting any sky and turning off the heliocentric velocity correction. From this cube we extracted the same spaxels corresponding to the position of our stars and co-added them, in the same fashion as was done for the science spectra. 
These sky spectra at the positions of our sources are then fitted with a model of the brightest OH sky-lines in that wavelength range. We brought the model OH spectra to the same resolution as the MUSE spectra. The precise restframe wavelengths of the OH line model was taken from the UVES sky emission catalog (\citealt{Hanuschik2003}). An example of the combined model of 13 OH lines is shown in blue in the top panel of Fig. \ref{fig:sky}. As MUSE spectra are undersampled with a FWHM of 2.4\,$\AA$ and a sampling of 1.25\,$\AA$/pix of the spectral axis, our model includes several sky lines, which are then simultaneously fitted to the spectra, in order to be not affected by the undersampling of a single line. Thus the only free parameter of the final fit is the wavelength shift. 

The velocity calibration of the MUSE spectra can vary from spaxel-to-spaxel across the entire field of view of the instrument. This variance of the sky line velocities is plotted in the bottom panel in Fig. \ref{fig:sky}. It illustrates why we cannot assume one absolute velocity calibration for all our stars. The analysis of the separate spaxels of the sky cube has been done in the same way as outlined above, shifting the sky model in wavelength direction. Due to the variance across the FOV we co-add the spaxels at the position of our sources and then derive the velocity calibration from those spectra. The velocity correction factors for the stars have a mean of $<v>=0.8835 \rm \,km\,s^{-1}$. 

Above we outlined how we estimated the velocity correction factor. The pattern across the field is of systematic nature caused by applying a wavelength solution taken during day calibrations at different times/temperatures of the instrument. The measurement and application of the velocity correction factor assumes implicitly that the original wavelength solution has no uncertainty in the first place. However, this is not the case and there are always some deviations in the wavelength solution, which we need to determine from the arc lamp exposure itself. This is the final systematic velocity error since we took the velocity offsets into account that occur when applying the wavelength solution to the science data taken at different conditions.

\begin{figure}
	\includegraphics[width=\columnwidth]{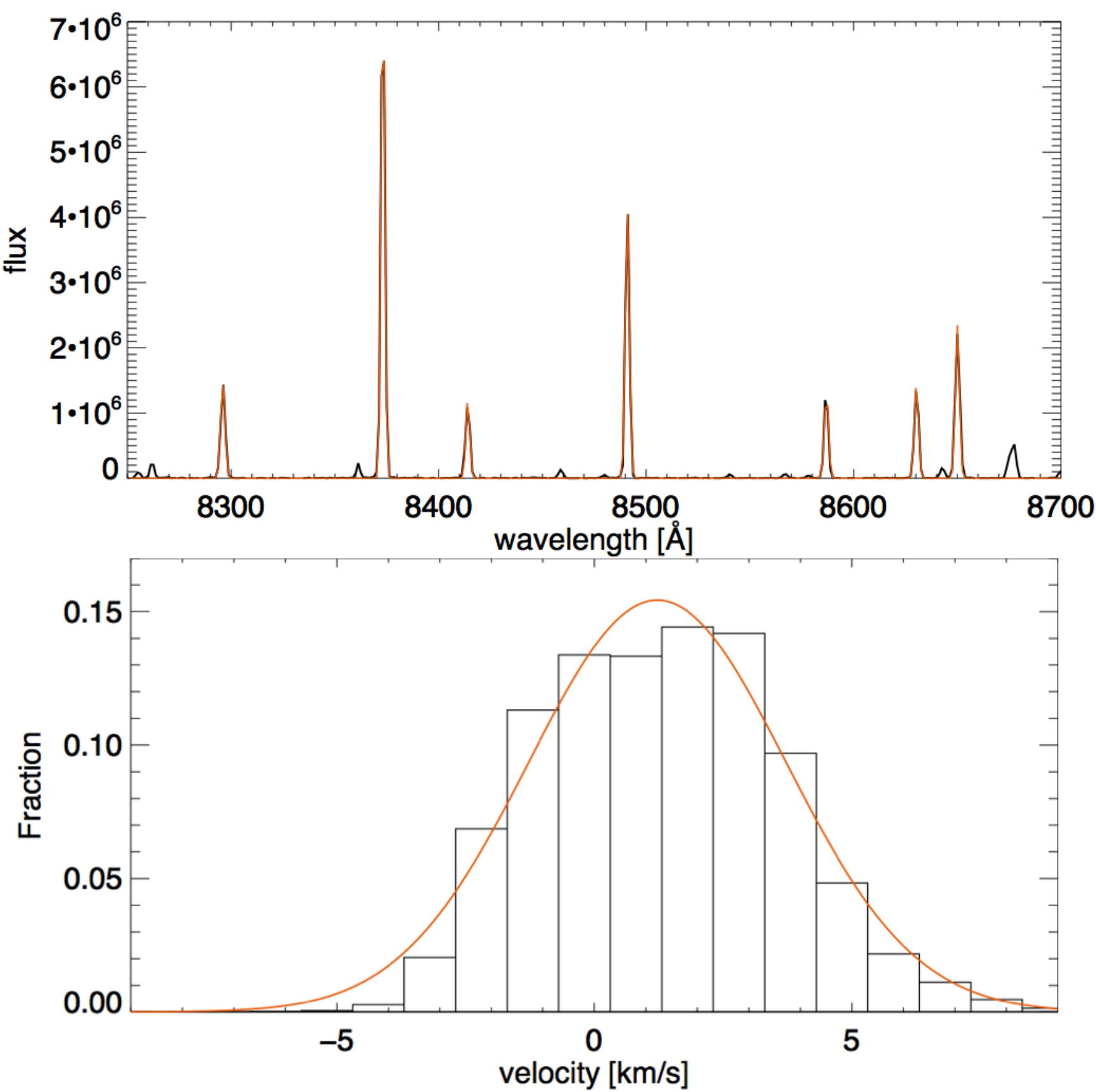}
    \caption{Top panel: The spectrum of the neon arc lamp in the Ca\,II triplet wavelength region, for one example spaxel. The fitted arc line model is shown in red.  Bottom panel: The histogram of the arc lamp velocity measurements in each of the spaxels of the MUSE arc lamp cube. Overplotted in red is the best fit Gaussian model with a width of $\sigma = 2.27 \rm \,km\,s^{-1}$ }
    \label{fig:arc}
\end{figure}

To constrain the systematic errors we used a MUSE observation of a neon arc lamp,  for which we applied a self-calibration by reducing  it as a science frame. No radial velocity offset needs to be taken into account since calibration and data are matched in time and conditions. This reduced arc lamp frame was then analysed using a similar method which was used to measure the final radial velocities. We use 7 lines in the neon spectrum in the same wavelength range as the NIR Ca II triplet (see top panel Fig. \ref{fig:arc}). This neon model is then fitted to the spectra of each of the 314$\times$304 spaxels in the MUSE cube. An example for one such fit is given in the top panel of Fig. \ref{fig:arc}. The histogram of this procedure is show in the bottom panel of Fig. \ref{fig:arc}. We fitted a Gaussian distribution to the histogram. The sigma of the Gaussian is  $\sigma = 2.27 \rm \,km\,s^{-1}$. This width of the distribution of the systematic velocity error is the intrinsic accuracy of the wavelength solution and also represents the uncertainty for the velocity correction factor. This error is therefore our final instrumental velocity uncertainty which is quadratically added to the error deduced from the signal-to-noise of the spectra.
%(\ref{fig:}). 

\subsection{Extraction of stellar spectra}
As a first step we need to extract the spectra of each star from the three dimensional cube for further analysis. Within a crowded field we have to locate the sources and deblend them if they overlap. To extract the spectra of a point source in an IFU we have to select the spaxels that correspond to a point source and co-add them into a one dimensional spectrum. To do this in an optimal way we have to take into account that the point spread function (PSF) changes with wavelength and is smaller at larger wavelengths. 

For that we use the \textsc{PampelMuse} software package (\citealt{Kamann2013}). This software is optimised to analyse integral-field spectroscopy data of stellar fields, by modelling the wavelength dependence of the PSF and then extracting and co-adding spaxels into one dimensional spectra.
To run \textsc{PampelMuse} an input catalogue of source positions and an estimate of their initial magnitudes is required, for which we used the photometry of \cite{Belokurov2014}. As a first step, \textsc{PampelMuse} identifies sources from the initial catalogue for which it can extract spectra and cross correlates their catalogue positions with the most likely positions within the MUSE field-of-view. The results of this correlation are displayed interactively, and after investigation the user can redo or accept the positions of the detected point sources.

After a successful position cross correlation, the pipeline continues with the actual source selection. Only those sources above the limiting photometric magnitude, with a S/N>3 and a minimum separation of 0.3\,FWHM widths from other bright sources are extracted as single sources. The faint sources that are below the S/N $\sim$ 3 limit are also extracted, and co-added into one combined "unresolved" spectrum.

The extraction of accurate point source spectra from IFS data cubes including the wavelength dependent size is done by the CUBEFIT routine in \textsc{PampelMuse}. It fits an analytic Gaussian or Moffat profile to each wavelength slice of the data cube for the brightest stars in the field of view. As a Gaussian PSF profile can underestimate the wings of the PSF, we chose a Moffat profile for our modelling.

\begin{figure}
	\includegraphics[width=\columnwidth]{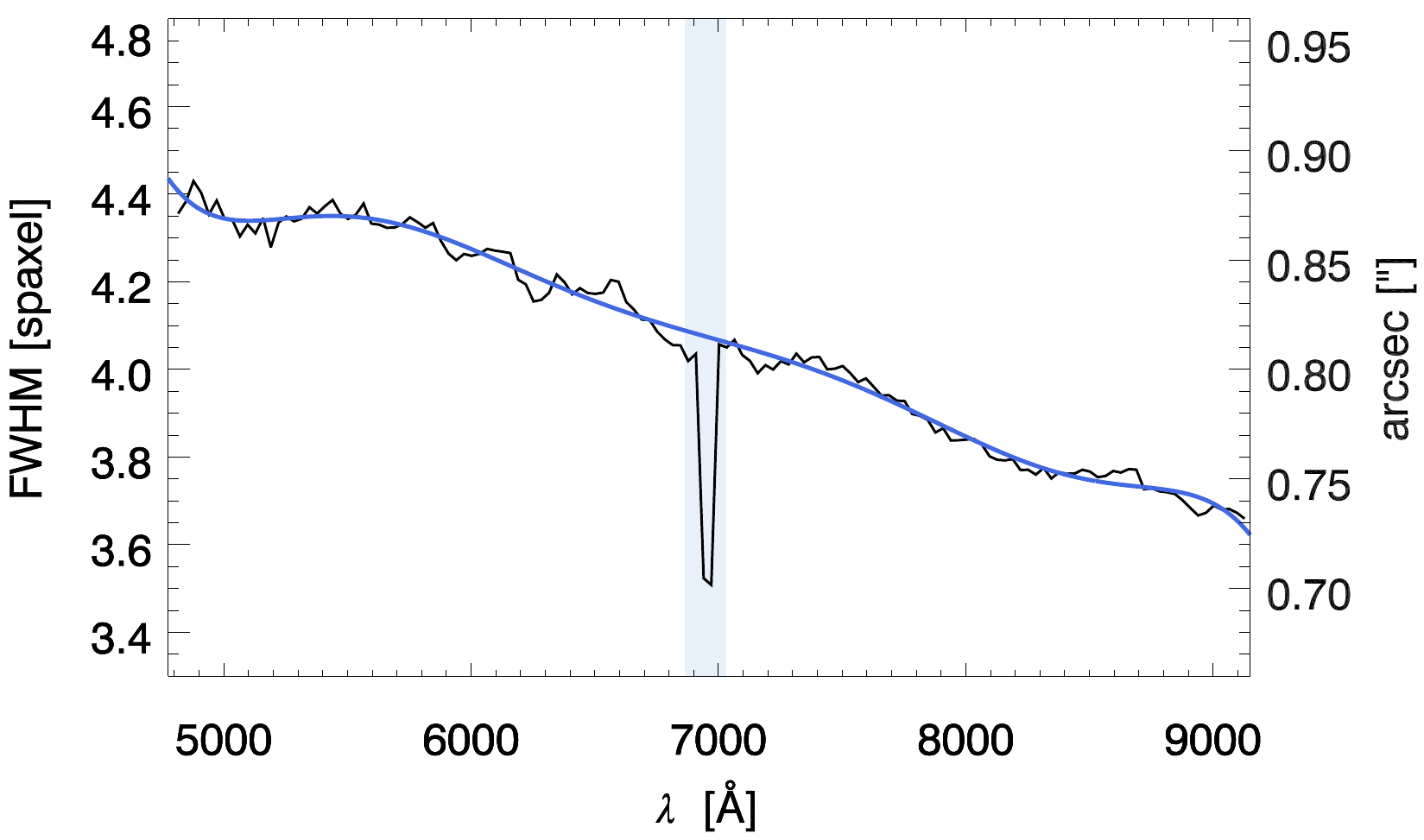}
    \caption{FWHM of the point spread function for each wavelength slice in the MUSE cube. The FWHM value was determined using a Moffat fit to the brightest stars in the field as determined by \textsc{PampelMuse}. In blue a Hermite Polynom of 7-th order fitted to the data is shown. The shaded region is an area of strong telluric absorption and was excluded from the fit. }
    \label{fig:PSF}
\end{figure}

The wavelength dependent FWHM of the PSF at each slice of the wavelength range is plotted in Fig. \ref{fig:PSF} as a black line. It was then fitted with a Hermite polynomial of 7th order, which is plotted in blue in Fig \ref{fig:PSF}. The blue shaded wavelength region is a strong telluric absorption band. It was excluded from the polynomial fit. In a last step \textsc{PampelMuse} stores each extracted one dimensional spectrum and the corresponding noise spectrum in a separate fits file that can be further analysed. We extracted a total of 41 spectra above the S/N>3 cut-off and one co-added spectrum of the unresolved sources. The position, magnitude, and colour of these point sources are shown in table \ref{tab:dat}. We extracted spectra in a magnitude range from $i=17.3$\, mag down to $i=22.1$\, mag.

\subsection{Radial velocity measurements}
We measure the radial velocities using the three strong and prominent absorption lines of the Ca\,II triplet in the NIR part of the spectra. These lines at 8498, 8542, and 8662\,\AA \,are excellent for determining the velocity of our stars, as they are strong features that are well resolved by medium resolution spectroscopy. 

We measure the most likely velocities of our stars by using a model Ca\,II triplet spectrum at rest wavelength, which is then fit to the observed spectrum by shifting in radial velocity. The model Ca\,II spectrum was constructed using a MILES Ca\,II spectrum (\citealt{Cenarro2001}) with the properties $T_{\rm eff}=$4750\,K, log(g)=1.25, [Fe/H]=$-$1.7\,dex with a spectral sampling of 1.25\,\AA, and a spectral resolution FWHM of 2.4\,\AA, similar to the MUSE instrumental values. We fitted the full Ca\,II MILES library to the spectrum of star 2, to determine a best fit model. The stellar parameters of the best fit model comparable well to those found for Crater stars in \citet{Bonifacio2015}. The final model spectrum was constructed by fitting three Gaussian absorption lines to the original MILES spectra simultaneously, while the centre of the lines was fixed to the precise velocity of the Ca\,II features.

\begin{figure*}
	\includegraphics[width=14.5cm]{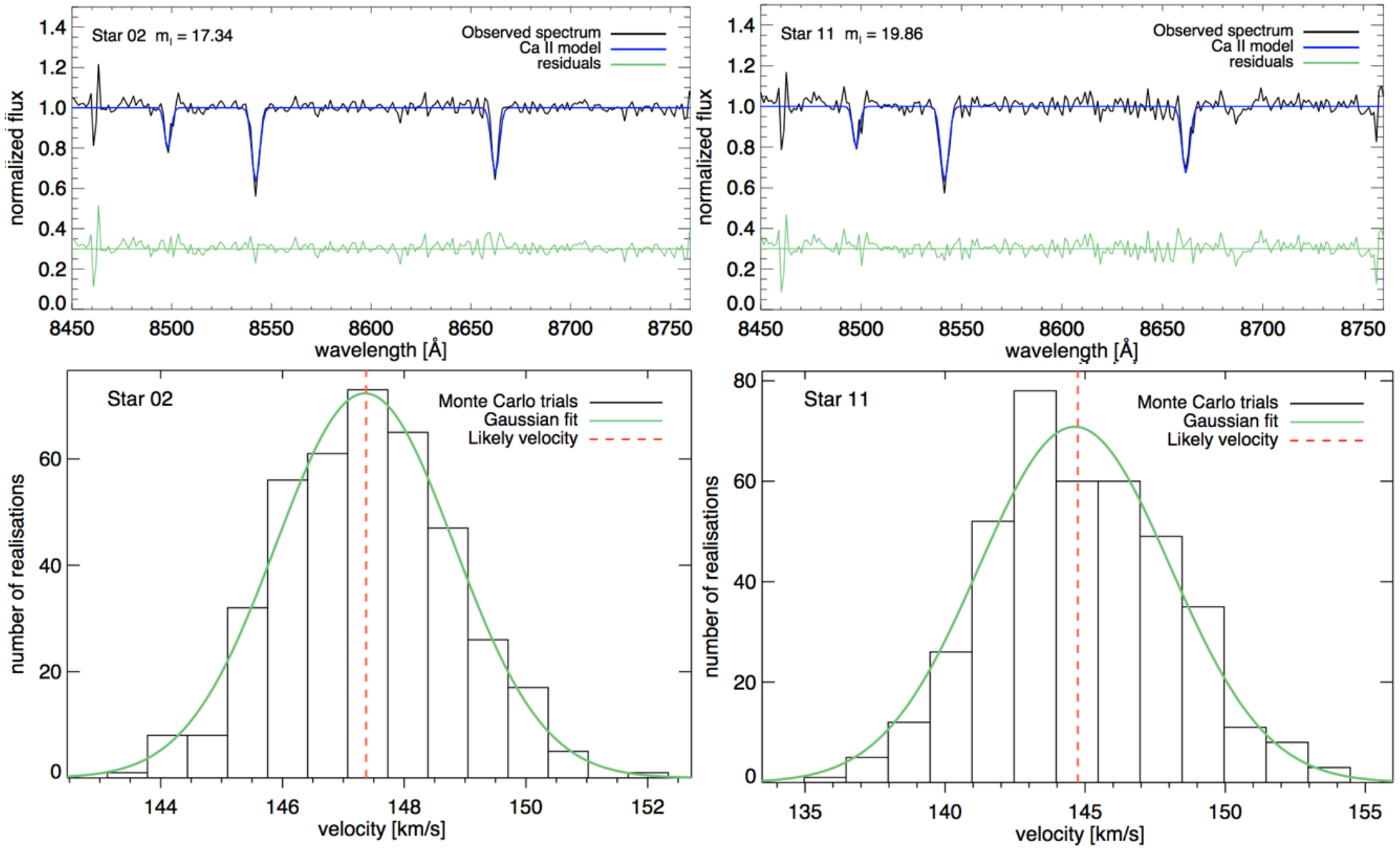}
    \caption{\textbf{Top left}: Continuum normalised spectrum of the Ca II region of star 2. The velocity shifted Ca II template spectrum is shown in blue, and in green the residuals of the fit are shown (offset by 0.3 for visibility). \textbf{Top right}: Same as the left panel but for star 11 which is 2.5 magnitudes fainter than star 2. \textbf{Bottom left}: The black velocity histogram shows the realisation of 400 Monte Carlo trials for star 2, when we added noise to the spectrum and then refitted the template. The green line is the best fit Gaussian, and the red dashed vertical line marks the most likely velocity, derived from the fit to the original spectrum. \textbf{Bottom right}: Same as the left panel but for star 11. }
    \label{fig:vel_star2}
\end{figure*}

To have robust estimates of the velocity errors introduced by the noise of the spectrum, we ran a Monte Carlo procedure where each stellar spectrum is resampled 400 times using the uncertainty of the spectrum. These resampled spectra are then each again fitted with the model Ca triplet spectrum. The distribution of the determined radial velocities of each star was then fitted with a Gaussian function to determine its 1-$\sigma$ errorbar. An example of such a Gaussian function for star number 2 is shown in figure \ref{fig:vel_star2}. The spectra were already corrected to the heliocentric rest frame by the reduction pipeline, and they were continuum normalised by dividing them with the best-fit third degree polynomial. The polynomial fit was carried out for each spectrum in the wavelength range between 8350\,\AA \,and 8900\,\AA. We applied a 2.5\,$\sigma$ clipping to our spectra in order to identify those wavelength regions dominated by continuum emission (as opposed to absorption lines). We then approximated the continuum by fitting a polynomial.

The derived noise uncertainty takes into account the error introduced by the noise of the spectrum, but not the error introduced by the systematics of the instrument, the undersampling or the local fitting process. The precision to which we can determine velocities with MUSE was determined in section \ref{sec:abs}. Using an arc lamp frame we found that the error introduced by the systematics, and thus the maximum precision of velocities derived from our MUSE data, is 2.27$\rm \,km\,s^{-1}$. As the error of the noise and the systematic error are independent from each other, we quadratically add the systematics and S/N errors of each extracted spectrum to arrive at a total uncertainty estimate.
The typical uncertainties of our velocities are in the range of 2.3 - 5 $\rm \,km\,s^{-1}$ for the brightest stars with $m_{\rm I}<$20.0 and between 5 - 15$\rm \,km\,s^{-1}$ for the fainter stars. A complete list of every star, its position, magnitude, colour and radial velocity is given in Table \ref{tab:dat} in the appendix.

  \begin{figure}
	\includegraphics[width=\columnwidth]{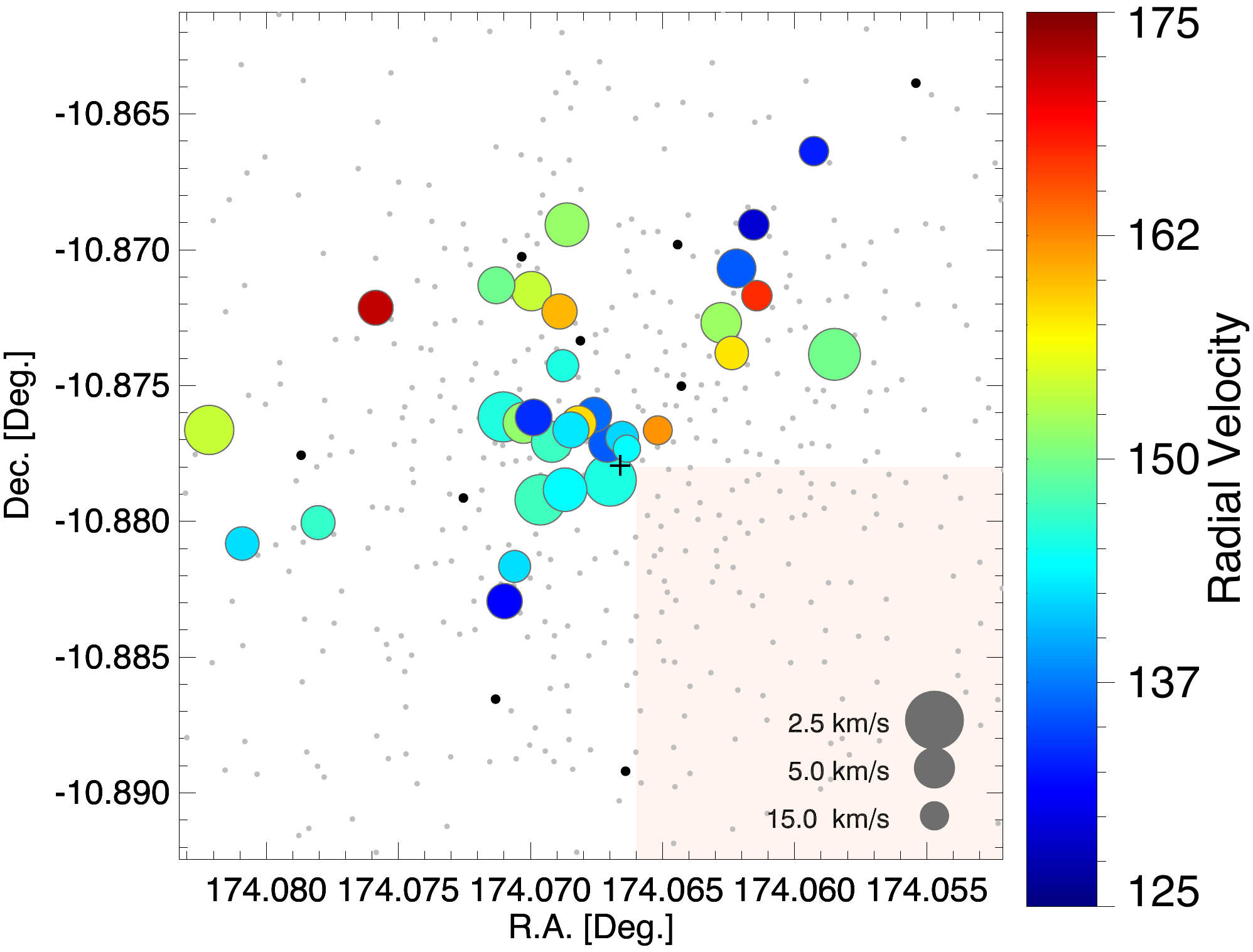}
    \caption{Position distribution of Crater stars with a measured velocity. The coloured circles represent the extracted stars, colour coded according to their radial velocity between 125 and 175\,km\,$s^{-1}$. Stars that have a measured velocity below or above the colour coding limits are shown as black points. The larger the size of the coloured circles, the lower the derived uncertainty of the velocity measurement. The small grey dots represent all other stars in the Crater region between $17.0<I_{mag}<26.0$ from \citet{Belokurov2014}. The shaded region marks the area that was not observed with MUSE }
    \label{fig:pos_vel}
\end{figure}
In Figure \ref{fig:pos_vel} the two dimensional position distribution of the stellar velocities is shown. Stars with a velocity between 125 and 175\,km\,$s^{-1}$ are colour coded according to their velocity. Stars with measured velocities outside this velocity range are plotted in dark black. Photometric measurements of stars in the FOV of Crater from the catalogue of \citet{Belokurov2014} in the magnitude range $17.0<I_{\rm mag}<26.0$ are shown as grey dots. The size of the coloured circles is a representation of their velocity uncertainty. One visible feature in the spatial distribution of the stars is the agglomeration of many stars close to the centre of Crater (marked by the black plus sign) with velocities close to the systemic velocity of Crater. It is notable that stars located farther away from Crater show larger differences to the systemic velocity, but their measurement errors are also larger. 

\subsection{Comparison to other velocity measurements of Crater stars}

 \begin{figure}
	\includegraphics[width=\columnwidth]{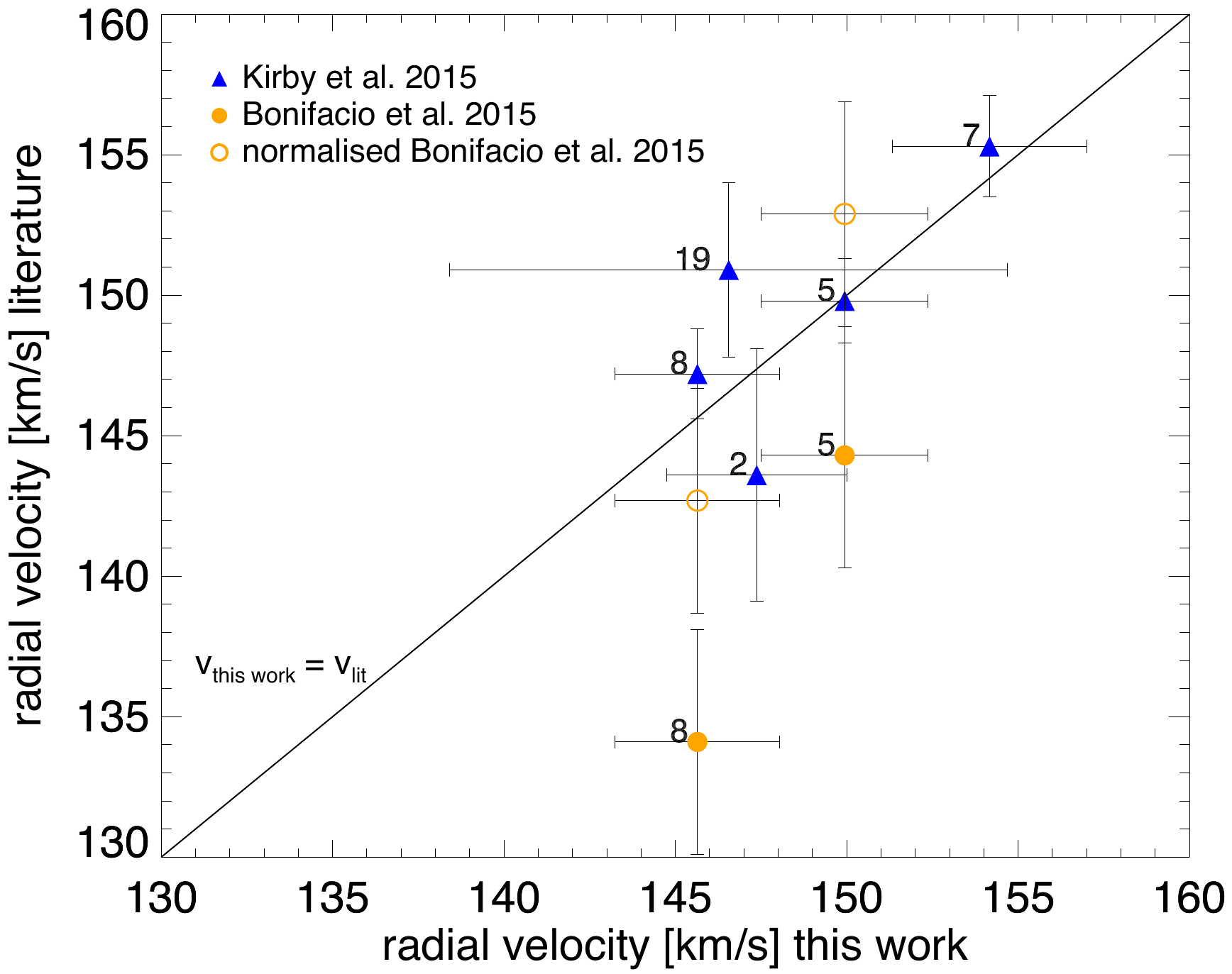}
    \caption{Comparison of stellar radial velocities with existing literature samples. On the x-axis our measurements are denoted and on the y-axis those from the literature. Blue triangles represent the values for the stars that were also measured in \citet{Kirby2015}, and filled orange circles denote the two stars measured in \citet{Bonifacio2015}. The numbers next to the data points are the identification numbers from our sample (see Table \ref{tab:dat} in the appendix). As there might be a systematic velocity offset in the Bonifacio sample, we also plot the open orange circles, which are the shifted Bonifacio velocities. They are shifted in such a way that they have the same mean velocities as stars 5 and 8 from our dataset. Stars 5 and 8 have velocity measurements in all three datasets.}
    \label{fig:lit}
\end{figure}
In this section we compare our velocity measurements for Crater stars with the two existing samples in the literature. In Fig. \ref{fig:lit} the radial velocities from the literature are plotted against the values we measure in this work. In total, 5 member stars from our sample are also present in \citet{Kirby2015}.  Both stars that were measured in \citet{Bonifacio2015} are present in all three samples. As is easily visible from the plot, our measurements are in overall very good agreement with the values given in the \citealt{Kirby2015} sample. Interestingly, for star 7, which is the blue star with ambiguous membership, the independent measurement and our value are in excellent agreement. Therefore it is likely that the radial velocity of this star is genuinely high. 

The two original velocities from the \citealt{Bonifacio2015} sample (filled orange circles Fig. \ref{fig:lit}) are outliers compared to our measurements. They are systematically lower than our values as well as those from \citealt{Kirby2015}. Our velocity of star 5, which is in all three samples, agrees very well with the \citealt{Kirby2015} value, but not with the \citealt{Bonifacio2015} value, which is significantly lower. This is also true for star 8 from of our sample which was also measured by both previous spectroscopic studies. While the \citealt{Kirby2015} value is consistent with ours, the \citealt{Bonifacio2015} value is a $3-\,\sigma$ outlier to both measurements. The radial velocities of stars that appear in our sample and the literature are summarised in table \ref{tab:dat}.
\begin{table*} 

\caption{Stellar radial velocities compared to those in the literature.}{} % title of Table 

\vspace{5 mm}

\centering      % used for centering table 
\begin{tabular}{c c c c c c c c c c}  
\hline\hline    
	&	velocity &	Index  & velocity & Index  & velocity \\
 & this work  & Kirby et al. & Kirby & Bonifacio et al.&  Bonifacio et al.  \\
 &(km s$^{-1}$) &  & (km s$^{-1}$) & & (km s$^{-1}$) \\
\hline 
      2 &        147.37 $\pm$        2.63 & 1710  & 143.6 $\pm$ 4.5 & - & - \\
       5 &       149.94 $\pm$        2.43 & 420  &149.8 $\pm$ 1.5 & J113615-105227 & 144.3 $\pm$ 4.0 \\
       7 &  	154.17 $\pm$        2.83 & 399  & 155.3 $\pm$ 1.8 & - & - \\
       8 &    145.64 $\pm$        2.40 & 93  & 147.2 $\pm$ 1.6 & J113615-105244 & 134.1 $\pm$ 4.0 \\
       19 &    146.56 $\pm$        8.14 & 1684  & 150.9 $\pm$ 3.1 & - & - \\

 \hline
\\

\end{tabular} 

\label{data}  % is used to refer this table in the text 

\begin{flushleft} This table lists the identifier of our work, the RA and DEC, the radial velocity from our sample and its errors. In column 3 the identifier from \citet{Kirby2015} is listed, and column 4 lists their velocities with errors. Column 5 and 6 list the respective values for the two stars from the \citet{Bonifacio2015} sample.
\end{flushleft}

\end{table*}

The large discrepancy of the two Bonifacio et al. values could also be due to a systematic velocity offset compared to our sample. Thus we renormalised the average velocity of both stars to the average velocity of those stars in our sample (open orange circles in  Fig. \ref{fig:lit}). While the corrected velocities are now less significant outliers, the velocity difference between star 5 and 8 of $10\rm \,km\,s^{-1}$ remains much larger than the difference of $4\rm \,km\,s^{-1}$ in our sample and $2.5 \rm \,km\,s^{-1}$ in the Kirby sample.

We also investigated when these observations were taken to determine if this could be radial velocity variations due to a binary star. Our observations were taken on March 13th 2015, the ones of Kirby on March 23rd 2015 and the ones of Bonifacio on March 22nd 2015 (as listed in the ESO Archive). Therefore we can exclude a real radial velocity variation, as the time delay between the Kirby et al. and Bonifacio et al. observation is a single day, and the difference to ours are 9 days. A binary is expected to have radial velocity variations on the order of a year. Thus we conclude that the most likely velocity of star 8 is consistent with the systemic velocity of Crater, and it is not a velocity outlier. Hence we suggest that the 95\% confidence level lower velocity dispersion limit of $\sigma>3.9\rm \,km\,s^{-1}$ by Bonifacio et al. based on the velocities of two stars only is likely wrong. 

\section{Probability based analysis of membership}
\subsection{Membership probability}
\label{sec:prob}
To determine which stars are members of Crater we have to distinguish them from foreground or background contaminants using our velocity, spatial, and colour information. Instead of making hard cuts for velocity, position, and CMD values we decided to follow a probabilistic method to statistically assess which stars are likely members. We base our method on what has been described in \citet{Collins2013} who investigated radial velocity measurements for 18 dwarf galaxies in Andromeda. The probability of each star to be a member is given as:
\begin{equation}
  P_{\rm n}=P_{\rm vel}\cdot P_{\rm dis}
  \label{eq:prb}
\end{equation}
where $P_{\rm vel}$ denotes the probability of membership based on the radial velocity of each star, whereas $P_{\rm dis}$ is the term that denotes the probability of membership based on the distance to Crater's centre.
In contrast to what has been done in \citet{Collins2013} we do not include a term $P_{\rm CMD}$ which penalises the probability of stars that are far from the red giant branch of the cluster CMD. This is done to avoid that the potential blue member stars are penalised, as they do not lie on the RGB.

The $P_{\rm dis}$ term of our probability function is based on the known radial profile of Crater. We model the probability with a normalised Plummer profile with a half-light radius $r_{\rm h}=0.46'$ (\citealt{Weisz2015}).
The posterior probability term $P_{\rm dis}$ can be written as:

\begin{equation}
P_{\rm dis} = \frac{1}{\pi \cdot r_{\rm h}^{2}[ 1+(r/r_{\rm h})^{2}]^{2}}
\end{equation}

\begin{figure}
	\includegraphics[width=\columnwidth]{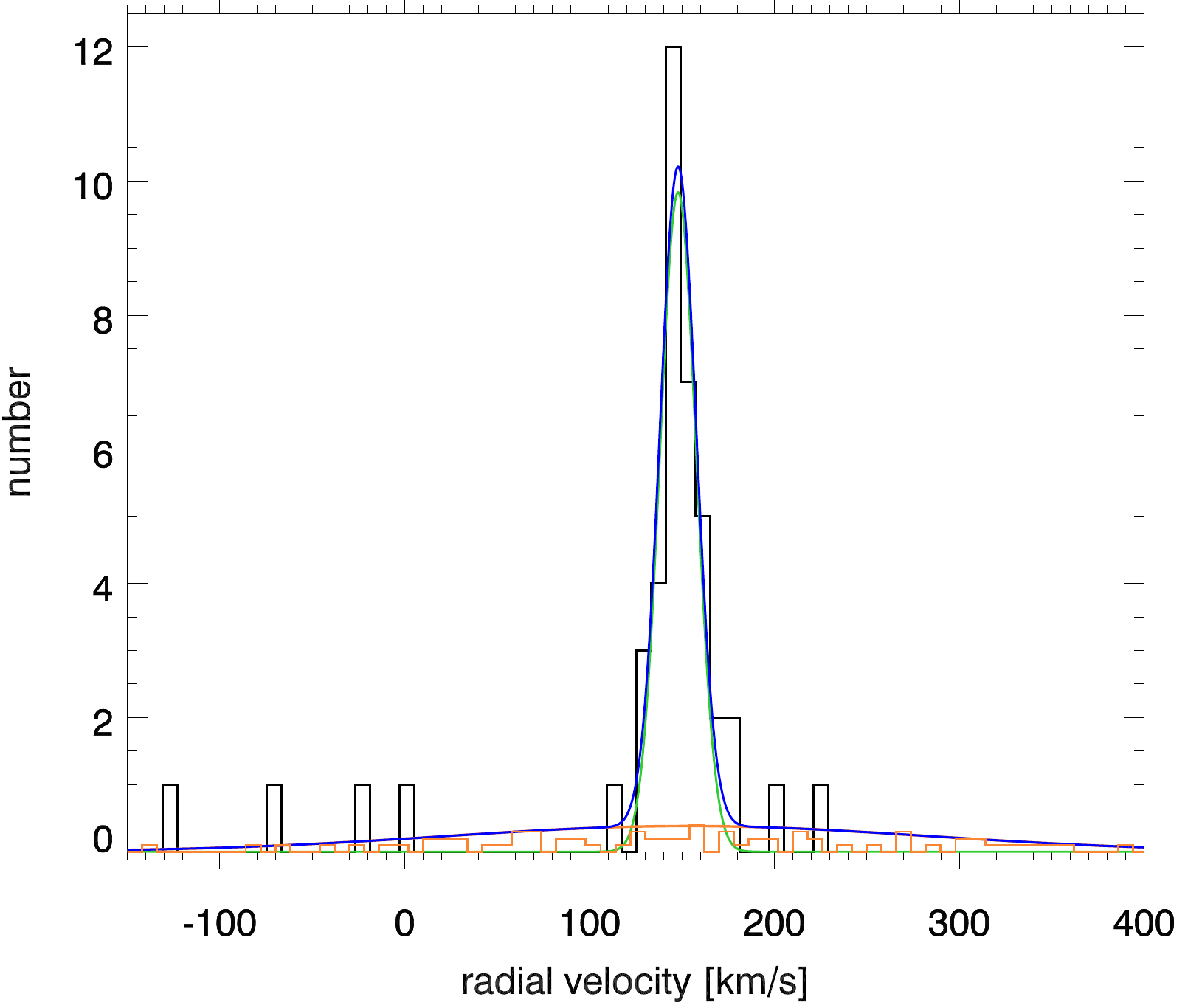}
    \caption{The velocity histogram for all observed stars in the MUSE field of view is shown in black. The strong peak of stars at v=148\,km/s is the systemic velocity of Crater, which clearly dominates our histogram. There are only few obvious foreground and background stars. The Besan\c{c}on foreground stars are shown as orange histogram. The simultaneous double Gaussian fit is shown as a blue curve and the results for the Crater population and foreground population are shown as green and orange Gaussian curves, respectively.}
    \label{fig:velocities}
\end{figure}

The probability term $P_{\rm vel}$ is determined by assessing how probable it is that a star is either part of the foreground Galactic contamination or a member of Crater. For this we use the velocity histogram of our measured velocities and model a Gaussian to  Craters velocity peak as well as to the foreground distribution. As we have only a very sparse sampling of the foreground stellar population, we use the Besan\c{c}on galaxy model (\citealt{Besancon2003}) to determine the expected shape of the velocity distribution of the Galactic foreground at the Galactic longitude and latitude of Crater. We retrieved the Galactic foreground velocity distribution from the Besan\c{c}on model for a FOV 10 times the size of the MUSE field, then normalise it to the same area. With this we ensure to have reliable number statistics on the foreground distribution. We choose to approximate the Besan\c{c}on foreground distribution with a broad Gaussian model. The central velocity and the width of the foreground distribution, as derived from the Besan\c{c}on model, are then used as fixed parameters of the foreground model when we decompose the observed data in foreground and Crater component. The systemic velocity of the Galactic contamination (red histogram in Fig. \ref{fig:velocities}) is found to be v$_{\rm r, gal}=153.1\rm \,km\,s^{-1}$ with a velocity dispersion of $\sigma_{\rm v, gal}=131.3\rm \,km\,s^{-1}$.

The measured velocities (blue histogram in Fig. \ref{fig:velocities}) were then fitted with the sum of two Gaussians, one for the Crater velocity and one for the Galactic foreground. Although the Galactic foreground has its central velocity close to the Crater peak velocity, 
 the level of Galactic contamination of 0.3 stars per velocity bin is small. The kinematic peak of Crater is clearly visible as a spike at v$_{\rm r}=148\rm \,km\,s^{-1}$ on top of the Galactic foreground.
The probability of each star to be a member of the Crater system is then given as:
\begin{equation}
P_{\rm n, crater}= a_{\rm 0, crater}\cdot \rm exp \bigg[ -\frac{1}{2} \, \frac{(v_{\rm n} - v_{\rm r, crater})^{2} }{\sigma_{\rm v, crater}^{2}+v_{\rm n, err}^{2}}\bigg]
\end{equation}
where the fitting of the double Gaussian has found the systemic velocity to be $v_{\rm r, crater}=147.9\rm \,km\,s^{-1}$ and a dispersion of $\sigma_{\rm v, crater}=4.47\rm \,km\,s^{-1}$. To calculate the probabilities of membership of the single stars we also take into account the velocity error of each single star adding the $v_{\rm n, err}^{2}$ term to the Gaussian. The galactic foreground component was modelled as: 
\begin{equation}
P_{\rm n, gal}= a_{\rm 0, gal}\cdot \rm exp \bigg[ -\frac{1}{2} \bigg( \frac{v_{\rm n} - v_{\rm r, gal} }{\sigma_{\rm v, gal}}\bigg)^{2} \bigg]
\end{equation}
where the systemic velocity of the foreground is given as $v_{\rm r, gal}=153.1\rm \,km\,s^{-1}$ with a broad width of the Gaussian distribution of $\sigma_{\rm v, gal}=131.3\rm \,km\,s^{-1}$ and a central height of the Gaussian of $a_{\rm 0, gal}=0.38$.

With the results from fitting a double Gaussian distribution to our data, each star can then be assigned a probability $P_{\rm vel}$ to be a member of Crater by dividing the probability to be a member of Crater by the sum of probabilities to belong to the galactic foreground or to Crater:

\begin{equation}
P_{\rm vel, n}=\frac{P_{\rm n, crater}}{P_{\rm n, crater}+P_{\rm n, gal}}
\end{equation}

\begin{figure}
	\includegraphics[width=\columnwidth]{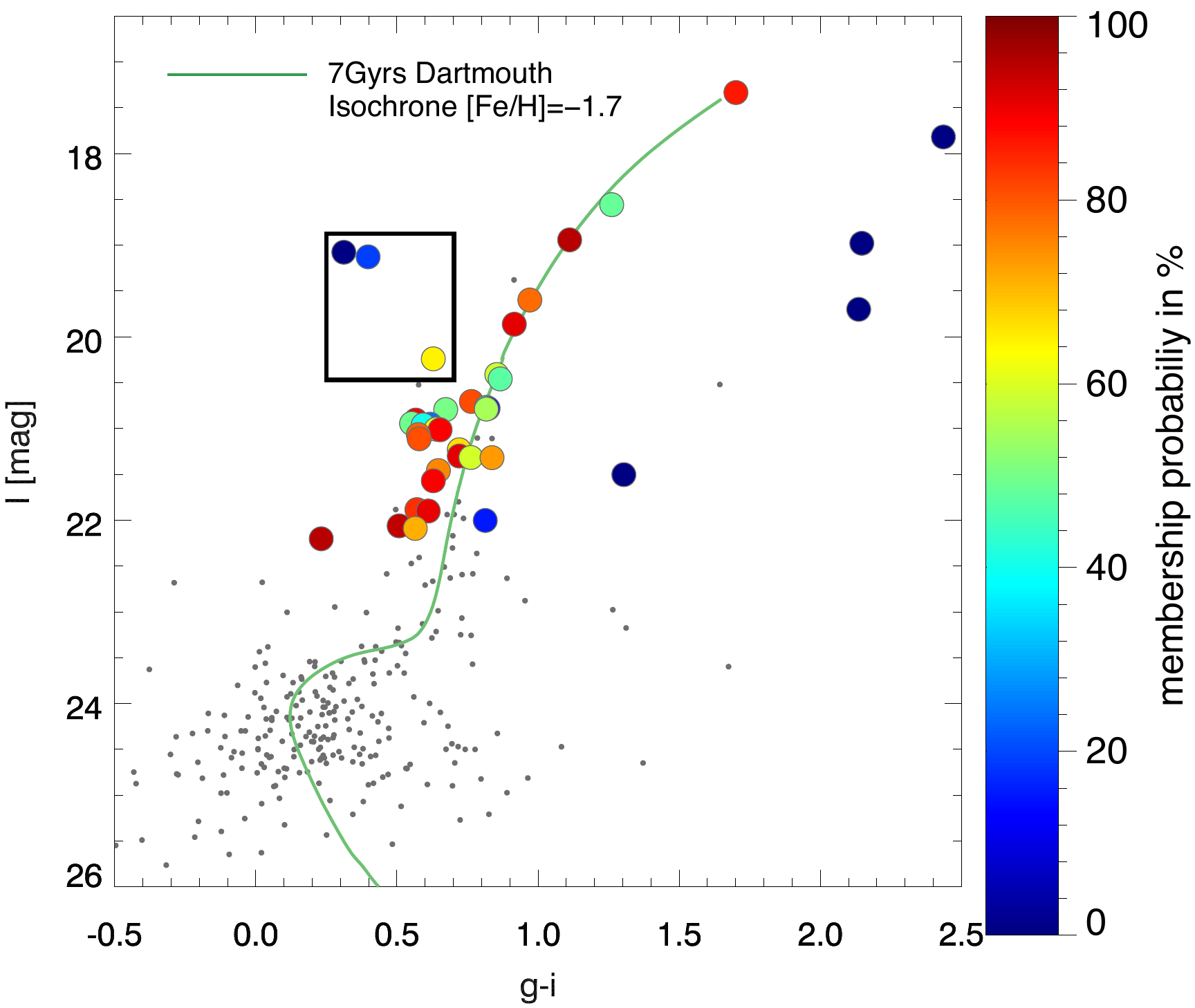}
    \caption{Crater CMD with stars for which we have a MUSE spectrum plotted as coloured points. All other stars are plotted as grey dots. The stars are colour coded according to their total membership probability. A Dartmouth isochrone with an age of 7\,Gyr and [Fe/H]=-1.7\,dex is plotted as green curve.}
    \label{fig:cmd}
\end{figure}

The assigned membership probabilities for all stars with a MUSE spectrum are displayed in Fig. \ref{fig:cmd} as colour coding in the colour magnitude diagram (CMD). The i-band magnitude and $g-i$ colour photometry are based on data from \citet{Belokurov2014}.

In the CMD it is clearly visible that the probability method works reasonably well, assigning very low probabilities to clear non-members in the CMD, e.g. stars offset from the RGB. The red giant branch (RGB) and the red clump are well populated with stars with a high membership probability. Notably, there are four stars above the red clump and blue-wards of the RGB, for which three have a spectrum and thus an assigned probability (CMD area marked with a black box). These have been noted as a potential intermediate age population in \citet{Belokurov2014}. The star marked in dark blue has $P_{\rm n}=10^{-17}$ due to its velocity of $v=222.3 \pm 3.67 \rm km\rm s^{-1}$  and is thus a clear non-member. For the two other stars the case is less clear. Right next to the non-member star in the box, star 7 is marked with a lighter shade of blue. It has a membership probability of $P_{\rm n}=0.20$. This relatively low probability is caused by its spatial position at one half-light radius from the centre and by the radial velocity of $ v_{r}=154.17\pm 2.8\rm \,km\,s^{-1}$ that is a $2\,\sigma$ outlier from the systemic velocity of Crater. Therefore, it seems possible that this star with a relative velocity of $\Delta v=6.5\rm \,km\,s^{-1}$ is bound to Crater. Considering its position at roughly one half-light radius, 52" from the centre, it is also possible that this is an unbound star that was recently stripped from the system. The third blue star with uncertain membership is star 13 marked by the yellow point within the black box in Fig. \ref{fig:cmd}. It has a probability of 64.9\,\% to be a member of Crater. With its radial velocity of $ v_{r}=151.6\pm 4.8\rm \,km\,s^{-1}$ it is consistent within $1\,\sigma$ of Crater's systemic velocity and its location at roughly half of the half-light radius makes it likely that it is a member star. Another notable feature is the blue faint star at $g-i=0.23$, located above the expected main sequence turn-off in the expected blue straggler region. Although this is the faintest star, for which we were still able to extract a spectrum, and thus its radial velocity measurement has a high uncertainty, it has a very high membership probability of P=95\,\% making it a likely member.

\subsection{Likelihood analysis}
To determine the systemic velocity and the velocity dispersion of Crater, we follow the Bayesian maximum likelihood technique as used in \citet{Collins2013} and established in \citet{Martin2007}. For this we use the posterior probability of each star being a member of Crater as a weight in the likelihood function. To determine which set of $[v_{r},\sigma_{v}]$ parameters maximises the likelihood function, we sample a fine grid in parameter space. The 1000$\times$1000 grid covers a range of 0-14$\rm \,km\,s^{-1}$ for the velocity dispersion and 135-160$\rm \,km\,s^{-1}$ for the systemic velocity. The log likelihood function that we are maximising can be written as:
  \begin{equation}
   \rm log(\mathcal{L}(v_{\rm r}, \sigma_{\rm v}))=-\sum_{n=0}^{N} \bigg[P_{\rm n} \rm log \,\sigma_{\rm tot} +\frac{1}{2}P_{\rm n} \left(\frac{v_{\rm r}-v_{\rm n}}{\sigma_{\rm tot}}\right)^{2}+P_{\rm n} \rm log(2\pi)\bigg]  
       \label{eq:ML}
   \end{equation}
 
Here N is given by the total number of stars for which we measured a radial velocity, $v_{\rm n}$ is the radial velocity of each star, and $P_{\rm n}$ is the posterior probability of the n-th star to be a member. The term $\sigma_{\rm tot}=\sqrt{\sigma_{\rm v}^{2}+v_{\rm n, err}^{2}} $ includes the velocity error for each star. With this way of measuring the intrinsic velocity and dispersion we can discriminate the intrinsic velocity dispersion of the object from the one introduced by the errors of the single velocity measurements. The errors of the velocity measurements are the quadratically added signal-to-noise errors and the systematic uncertainty of $2.27\rm \,km\,s^{-1}$.

\section{The kinematics of Crater}
\subsection{Velocity dispersion and M/L}
We applied the previously described maximum likelihood analysis to our sample stars, using the posterior probabilities from Sec. \ref{sec:prob} as weights for the individual stars. We applied a two dimensional grid of the systemic velocity $v_{\rm sys}$ and the velocity dispersion $\sigma_{\rm v}$ to determine the parameter set that maximises our likelihood function $\mathcal{L}(v_{\rm r}, \sigma_{\rm v})$.

\begin{figure*}
	\includegraphics[width=1.0\textwidth]{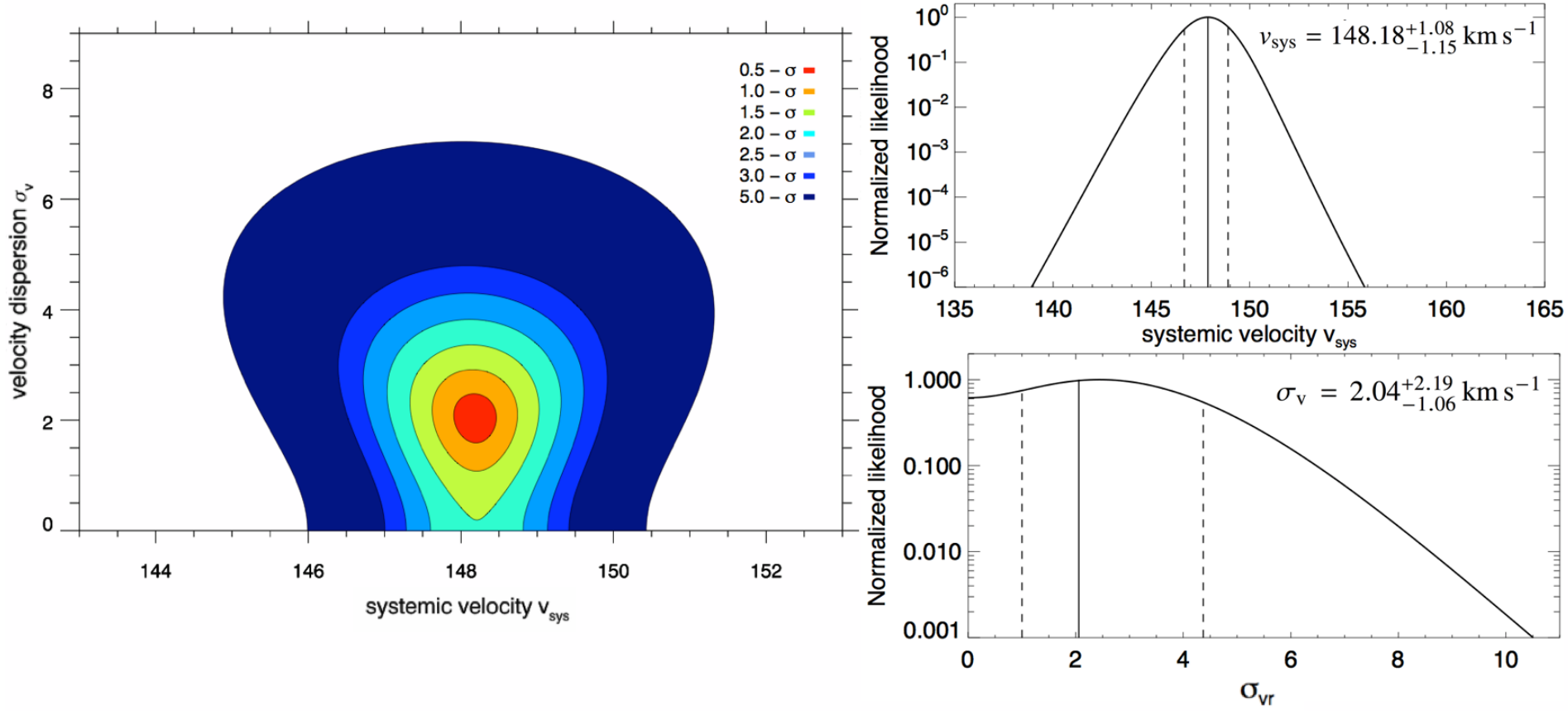}
    \caption{\textbf{Top left panel}: Two dimensional distribution of the normalised likelihood for the grid of systemic velocity and velocity dispersion. The colour coding of the likelihood ranges from 0.5-3.0\,$\sigma$ levels in steps of 0.5 and an additional colour level for $5\,\sigma$. \textbf{Top right panel}: The marginalised one dimensional likelihood distribution for the systemic velocity of Crater. The most likely value is noted with a solid line, whereas the 1.0\,$\sigma$ uncertainty levels are shown as dashed lines. \textbf{Lower right panel:} The marginalised one dimensional likelihood distribution of the velocity dispersion. The solid line denotes the most likely value and the dashed lines mark the 1-$\sigma$ levels.}
    \label{fig:dyn}
\end{figure*}

The two dimensional distribution of the normalised likelihood for the grid of $\sigma_{\rm v}$ and the systemic velocity $v_{\rm sys}$ is shown in the left panel of \ref{fig:dyn}. The coloured sigma contours are the likelihood levels corresponding to the sigma levels of the systemic velocity, which has an almost symmetric likelihood distribution. We use the mean of the lower and upper velocity error, which is 1.11$\rm \,km\,s^{-1}$ to trace the likelihood levels that correspond to the noted 0.5, 1.0, 1.5, 2.0, 3.0 and 5.0$\,\sigma$ levels. The explicit likelihood levels that are traced by this method are 0.9641, 0.8640, 0.7196, 0.5571, 0.4009, 0.2682 and 0.0258 in increasing order of sigma contours. The marginalised one dimensional likelihood distributions for both values are shown in the two right-handed panels. The most likely values are indicated with a vertical line, as well as their 1\,$\sigma$ uncertainties with dashed lines.

The set of values that maximises the likelihood is given as: $v_{\rm sys}=148.18^{\rm +1.07}_{\rm -1.15}\rm \,km\,s^{-1}$ and  $\sigma_{\rm v}=2.04^{\rm +2.19}_{\rm -1.06}  \rm \,km\,s^{-1}$. We are able to resolve the velocity dispersion with our full sample and a dispersion of 0 is excluded with a confidence of almost  $2\,\sigma$ at the lower end. Also we can exclude a velocity dispersion of $\sigma_{\rm v}>6\rm \,km\,s^{-1}$ with 2 sigma confidence. The larger uncertainty of the dispersion towards higher values is due to several stars that are velocity outliers and have medium or low probability to be members and thus their contribution to the likelihood function does not vanish. 

We tested how robustly our routine handles velocities with large uncertainties, of which we have several in our sample. Thus we reran the determination of the velocity dispersion, limiting our sample to stars with $v_{\rm err}<15\rm \,km\,s^{-1}$, $v_{\rm err}<10\rm \,km\,s^{-1}$ and $v_{\rm err}<7\rm \,km\,s^{-1}$. The systemic velocity and the dispersion of Crater change only marginally when excluding low S/N stars. The values are all compatible with the values for the full sample within their uncertainties. The derived velocity dispersions and the amount of included stars are:
\begin{itemize}
\item $\sigma_{\rm err<15km/s }=2.06^{\rm +2.2}_{\rm -1.06}  \rm \,km\,s^{-1}$ \quad  N=35
\item $\sigma_{\rm err<10km/s }=2.06^{\rm +2.1}_{\rm -1.05}  \rm \,km\,s^{-1}$ \quad N=29
\item $\sigma_{\rm err<7km/s }=2.16^{\rm +2.5}_{\rm -1.03}  \rm \,km\,s^{-1}$ \, \quad N=20
\end{itemize}
This shows that the determination of our systematic errors and the S/N errors using Monte Carlo resampling are a reliable estimate of our uncertainties, as the derived $v_{\rm sys}$ and $\sigma_{v}$  are stable when including or excluding low S/N stars.

We reran the same exercise for different membership probability cut-off limits. 
\begin{itemize}
\item $\sigma_{\rm p \geq 0.2 }=2.00^{\rm +2.3}_{\rm -1.03}  \rm \,km\,s^{-1}$ \quad  N=33
\item $\sigma_{\rm p \geq 0.5 }=1.43^{\rm +2.1}_{\rm -0.66}  \rm \,km\,s^{-1}$ \quad N=26
\end{itemize}
For the case where we limit our sample to stars with a membership probability of $\geq 0.5$ we resolve a smaller dispersion but one that is still consistent within the errors with our other samples.
From the spatial distributions of velocities in Fig. \ref{fig:pos_vel} it appears that the velocity dispersion of the central stars is lower than that of the outer stars. To test the isotropy of the velocity dispersion, we reran our Monte Carlo analysis for stars located within a radial distance of r$<$35\,arcsec. For the 24 stars in this sample we derive a likely systemic velocity of $v_{\rm sys}=147.17^{\rm +1.3}_{\rm -1.3}  \rm \,km\,s^{-1}$ and a velocity dispersion of $\sigma_{\rm v}=1.57^{\rm +3.66}_{\rm -0.54}  \rm \,km\,s^{-1}$. Thus the velocity dispersion is lower when only considering the central regions. Still, within the error bars, the values are largely consistent with the full sample values.

Assuming Crater is in dynamical equilibrium, we use the formula for deriving the dynamical mass of stellar dispersion supported systems of \citet{Wolf2010} to estimate its half-mass:
\begin{equation}
M_{1/2}=\frac{4}{G}\sigma_{v}^{2}\cdot R_{\rm e}
\end{equation}
R$_{\rm e}$ is the half-light radius of the system, $\sigma_{\rm v}$ the two dimensional velocity dispersion and G the gravitational constant. For $R_{\rm e}$ we use the value of 19.4\,pc given in \citet{Weisz2015}.
We derive a dynamical half-mass of Crater of $M_{\rm 1/2}=7.51^{+24.7}_{-5.8}\cdot10^{4}M_{\odot}$, assuming that it is a dispersion supported system in equilibrium. The mass-to-light ratio of Crater can then be calculated using the half-light luminosity $L_{\rm v}=8.8\cdot10^{3}L_{\odot}$ converted from the values given in \citet{Weisz2015}. We derive $M/L_{\rm V}=8.52^{+28.0}_{-6.5}\, M_{\odot}/L_{\odot}$.
For the second case, when restricting the measurement to the inner 35\,arcsec, we derive a mass of $M_{\rm 1/2}=4.45^{+44.9}_{-2.53}\cdot10^{4}M_{\odot}$ which translates to a mass-to-light ratio of $M/L_{\rm V}=5.05^{+50.9}_{-2.88}\, M_{\odot}/L_{\odot}$. 

The effects of dynamical equilibrium and spherical symmetry assumptions on the accuracy of the \citet{Wolf2010} mass estimator was examined recently by \citet{campbell2016}. They find that there exists no systematic biases, but an intrinsic 25\% 1-$\sigma$ scatter in the masses determined with this estimator. This effect is mainly due to the unknown 3D distribution of the stellar mass. This intrinsic scatter can increase the errors on our measurement further.

We use population synthesis models to compare predicted mass-to-light ratios of a purely baryonic stellar system to the measured M/L of Crater. These models need the age and metallicity of the cluster as input parameters. We adopt the best fit age of 7.5\,Gyr and a metallicity of [M/H]=\,$-$1.66 as derived from HST photometry in \citet{Weisz2015}. For this age and metallicity the SSP model of \citet{Maraston2005} predicts M/L$_{\rm v}=2.2\, M_{\odot}/L_{\odot}$ for a Salpeter IMF and M/L$_{\rm v}=1.5\, M_{\odot}/L_{\odot}$ for a Kroupa IMF. Both values are clearly lower than what we derive for Crater. The Salpeter IMF prediction lies within the $1\,\sigma$ uncertainty of the measured Crater M/L of the full sample, which has a lower limit of 2.17. The Kroupa IMF prediction is a slightly larger outlier. Thus we conclude that the measured M/L ratio of Crater is consistent with M/L predictions for a purely stellar system. Even with the assumption of a Kroupa mass function, the stellar contribution is maximised.

Mass functions flatter than Kroupa or Salpeter with correspondingly lower M/L-values are still possible (\citealt{Paust2010, Sollima2012}). Of course, all of this analysis is based on the assumption of dynamical equilibrium, which we will investigate further in the next section. If we just use the sample of stars within 35\,arcsec, the prediction for the M/L becomes a bit lower with 5.05, and its lower $1\,\sigma$ level is M/L$_{\rm V}=1.97\, M_{\odot}/L_{\odot}$, which is in good agreement with a stellar system following a Salpeter IMF. The upper limit for the dispersion and thus the M/L$_{\rm V}$ ratio of Crater is less tightly constrained and reflects the existence of several velocity outlier stars (such as star 7 with $154\rm \,km\,s^{-1}$) with a non-negligible membership probability. 

\subsection{Radial distribution of stellar velocities, cluster rotation and velocity gradient}
As the presence of rotational support, tidal distortions or binary stars can inflate the velocity dispersion significantly we test our sample as to whether we can find evidence for such a behaviour.

\begin{figure}
 \centering
	\includegraphics[width=1.0\columnwidth]{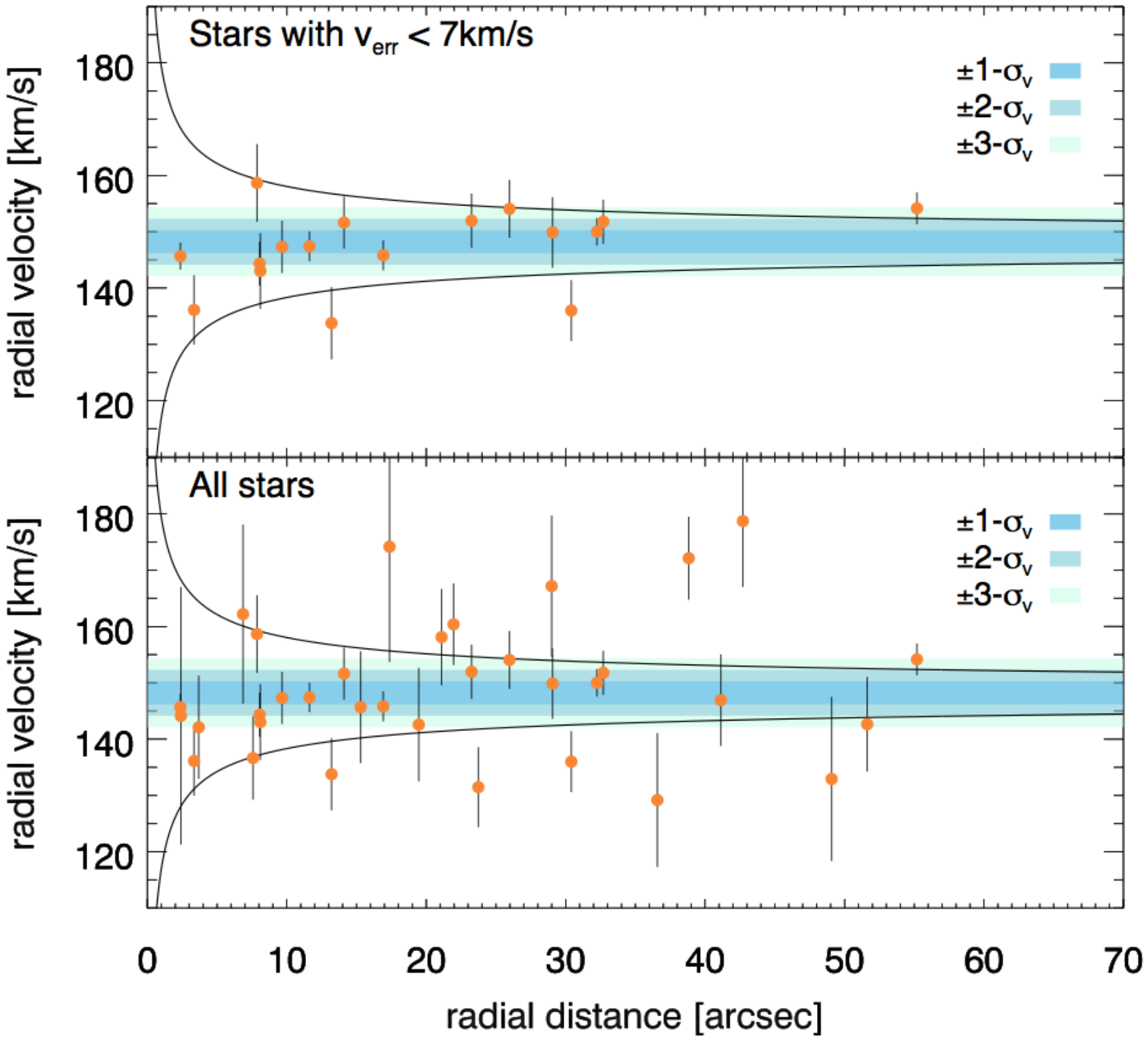}
    \caption{Stars with radial velocities as function of their distance to the centre of Crater. Shaded regions correspond to the velocity areas within $\pm 1\,\sigma_{\rm v}$, $\pm 2\,\sigma_{\rm v}$ and $\pm 3\,\sigma_{\rm v}$ of the systemic velocity of Crater. The top panel shows only velocities with $v_{\rm err}<7  \rm \,km\,s^{-1}$ and the bottom panel shows all stars with velocities in the displayed range. The black lines in both panels represent the radial dependence of the escape velocity of the system if we assume a total dynamical mass of $M_{\rm tot}=1.5\cdot10^{5}M_{\odot}$. }
    \label{fig:radial}
\end{figure}

The distribution of stellar radial velocities as a function of distance to Crater is shown in Fig. \ref{fig:radial}. In the top panel, only stars in the FOV that have a velocity uncertainty of less than $7\rm \,km\,s^{-1}$ are included, in the bottom panel all stars for which we measure velocities are shown. In the top panel, it is visible that most stars are consistent within the $\pm\,1\,\sigma_{\rm v}$ shaded area (darkest shade of blue) of the measured intrinsic dispersion of 2.05$\rm \,km\,s^{-1}$. There appears to be a subset of stars at larger and lower velocities whose errorbars are not consistent with the $3\,\sigma$ region of the intrinsic velocity dispersion. The star at the largest radial distance of 55\,arcsec is star 7, one of the potential blue stars of our dataset. It is within $2\,\sigma$ of the intrinsic dispersion of the system. The stars with good velocity measurements in the top panel of \ref{fig:radial} appear to follow a radial trend towards higher velocities for increasing distance to Crater. However, this behavior is not statistically significant, and we might be biased by the small number statistics of bright stars for which we have precise velocity measurements.

\begin{figure}
 \centering
	\includegraphics[width=1.0\columnwidth]{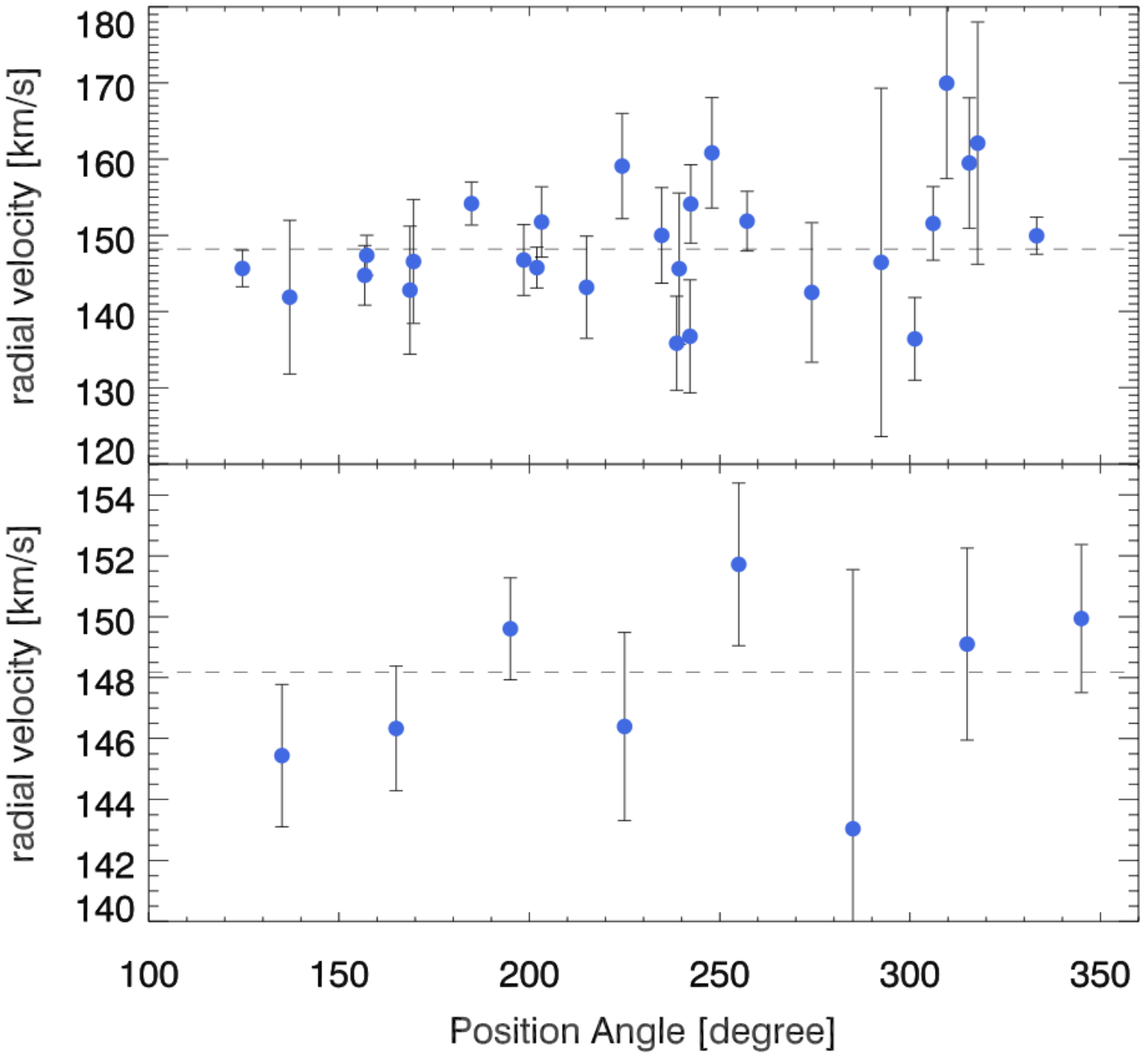}
    \caption{\textbf{Top panel:} The velocity of Crater stars as a function of the position angle (0 deg West, +90 deg South). Included are all stars with velocities between $130<v_{\rm rad}<175\rm \,km\,s^{-1} $. The dashed horizontal line in both panels indicates the systemic velocity of Crater. \textbf{Bottom panel:} The weighted mean of the radial velocities as a function of the position angle of the stars. The bin size is $30^{\circ}$. Note the different scaling of the y-axis in the two panels.}
    \label{fig:rotation}
\end{figure}

Our large sample of likely member stars enables us to test also for potential signatures of rotation in Crater. For this we plot the stars and their radial velocities as a function of their position angle in Fig. \ref{fig:rotation}. Here the position angle (PA) is defined as PA=0 in western direction and as +90 degrees towards the south. Thus, the first 90 degrees of the rotation plot are empty because they lie in the unobserved quadrant. In the top panel the position angle of all stars in the sample between $130<v_{\rm rad}<175\rm \,km\,s^{-1} $ are plotted. The plot indicates that there might be a radial velocity increase with position angle. To test if this increase is significant we used bins of $30^{\circ}$ and calculated the weighted mean radial velocity in each bin. The results are shown in the bottom panel of Fig. \ref{fig:rotation}.

The error-weighted mean velocities appear to depend on the position angle. As our data do not cover the full Crater position angles, and our velocities have relatively large errors, it is not possible to draw a firm conclusion on a rotational amplitude by fitting a sinusoidal curve to the data. Nevertheless, we see that the weighted mean of the bin at PA=$135^{\circ}$ is $v = 145.44\pm 2.34 \rm \,km\,s^{-1}$ and at a PA=$345^{\circ}$ it rises to $v = 149.94 \pm 2.43 \rm \,km\,s^{-1}$. Assuming these two values would represent the amplitude of a rotational support in Crater, the difference of  $4.5\rm \,km\,s^{-1}$ would imply an rotational amplitude of the order of $2.25 \rm \,km\,s^{-1}$. To determine if Crater is rotationally supported more and more precise radial velocities will be necessary. Alternatively, our data could also indicate that there is a velocity gradient from the south to the north, which would hint at previous tidal distortions of Crater's stellar population.

\section{Discussion}
\subsection{Dynamical state of Crater}

The dynamical mass-to-light ratio of Crater is consistent with predictions for baryonic globular cluster SSP models with a Salpeter IMF within its errorbars. Our derived total dynamical mass is $M_{\rm tot}=1.50^{+4.9}_{-1.2}\cdot 10^{\rm 5}M_{\odot}$ and translates into a mass-to-light ratio of $M/L_{\rm V}=8.52^{+28.0}_{-6.5}\, M_{\odot}/L_{\odot}$. This is consistent at the lower limit within $1\,\sigma$ with the M/L$_{\rm V}=2.2\, M_{\odot}/L_{\odot}$ predictions from SSP models of \citet{Maraston2005} for a stellar system of 7\,Gyr. 

One explanation for an elevated dynamical mass-to-light ratio can be anisotropies in the velocity dispersion. Anisotropies such as rotational support or velocity gradients can inflate the observed velocity dispersion and elevate the M/L ratio. The (though not significant) hints for rotational support in Crater, as well as the observed dependence of radial velocities on the distance to Crater, suggests that Crater might not be in dynamical equilibrium.

If we assume that Crater is not in dynamical equilibrium because it is tidally disturbed, the velocity outlier stars might have been recently stripped and are thus unbound stars that are still in the vicinity of Crater. In \citet{Kuepper2011} it was suggested that during the phase near the apocenter, tidal debris can be orbitally compressed and already unbound stars can appear to be close enough to be bound to the object. Thus, an object affected by tidal compression can appear dynamically hotter than it actually is, which will inflate the derived M/L ratios and can mimic the dynamical effects of a dark matter halo. 

Binary blue stragglers and other binary stars can inflate the velocity dispersion of clusters as they are observed at random times of their orbits (\citealt{Mcconn2010, Frank2012}). One faint blue star at I=22\,mag just above the main-sequence turn-off is a very likely member of Crater and located exactly where we would expect blue stragglers in the CMD. In addition its velocity is consistent with the systemic velocity of Crater. Its location at the very centre of the cluster at a distance of merely 6\,arcsec is consistent with predictions for blue stragglers to migrate towards the cluster centre over time (\citealt{Ferraro2012}).

In \cite{Milone2012} it was found that lower mass and less concentrated GCs have on average a higher binary fraction. This is attributed to the fact that dense stellar environments are more efficient in destroying binaries. As Crater has a low stellar content and is extended it should have a higher binary fraction than higher mass clusters. Assuming that Crater formed with its size and mass this means that we expect a high fraction of binaries in its stellar population that can then increase our velocity dispersion measurement. 

In \citet{Frank2012} the effect of binaries on the measured velocity dispersion was simulated for the low-mass GC Pal 4. They conclude that a high binary fraction in a low-mass clusters can result in estimating too high dynamical masses for a given cluster. They suggest that the velocity dispersion can be overestimated by 20\% assuming a high binary fraction within the cluster.

Crater's $M/L$ ratio is not in agreement with what we expect for a dwarf galaxy of this mass, having mass-to-light ratios of $M/L$>100. At face value, Craters $M/L$ ratio could nevertheless be interpreted as presence of dark matter. As we will discuss later, most arguments are in favour of Crater being a globular cluster, and the existence of dark matter in globular clusters has not been proven so far, although it was proposed several times in the past (e.g.: \citealt{Peebles1984,Padoan1997}).

Another explanation for the elevated dynamical mass of Crater might be that its dynamics does not follow Newton's laws but is non-Newtonian.
In the case of MOdified Newtonian Dynamics (MOND, \citealt{Milgrom1983}) one would expect a velocity dispersion of 2.2 km/s, which is in good agreement with our measured value. The MONDian velocity dispersion was calculated using eq. 6 in \citet{Baumgardt2005}, assuming that the cluster is in the deep MOND regime and that the internal acceleration dominates over the external one. We note, however, that other similarly remote clusters located in the low acceleration regime, like Pal\,14, Pal\,4, and NGC\,2419, do not show evidence for MONDian dynamics (\citealt{Jordi2009, Frank2012,Ibata2011}).

In \citet{Baumgardt2010} two distinct populations of outer halo Milky Way star clusters were identified based on to which extent they fill their Jacobi radius $r_{\rm J}$ in comparison to their half mass radius $r_{\rm h}$. On the one hand, there is a group of compact, tidally underfilling clusters with $r_{\rm h}/r_{\rm J}<0.05$. On the other hand, there exists a second population of tidally filling clusters $0.1<r_{\rm h}/r_{\rm J}<0.3$ that are likely in the stage of dissolution. Thus we can use this estimator to test if Crater is a dissolving GC. We use equation 1 from \citet{Baumgardt2010} to determine the Jacobi radius:
\begin{equation}
r_{\rm J}=\left ( \frac{GM_{\rm c}}{2\cdot V_{\rm G}^{2}} \right )^{1/3} R_{\rm GC}^{2/3}
\end{equation}
We use a dynamical total mass of $M_{\rm tot}=1.5\cdot 10^{5}\,M_{\odot}$ as the cluster mass, $V_{\rm G}=$220\,km/s as circular velocity of the Milky Way, and a Galactocentric distance for Crater of d=145\,kpc (taken from \citealt{Weisz2015}). Using a deprojected radius of $r_{\rm h}=1.33\cdot r_{\rm half light}$ as defined in \citet{Baumgardt2010}, we derive a ratio of $r_{\rm t}/r_{\rm J}=0.114$. This is consistent with a dissolving globular cluster in the tidally filling regime. Due to its unusual metallicity, age, and distance, it has been already speculated that Crater formed as GC of another dwarf galaxy that was then accreted onto the Milky Way. In this scenario Crater must have already been subjected to strong tidal forces during a pericenter passage in order to be efficiently stripped from its parent dwarf galaxy, and thus might currently be in the phase of dissolution.

\subsection{Is Crater a former member of a dwarf galaxy?}

It was noted by \citet{Belokurov2014} that Crater is aligned on a common great circle with Leo\,IV and Leo\,V, which also have comparable radial velocities. Thus a potential association of Crater with Leo\,IV and Leo\,V, with a common origin of these objects, has been suggested. Recently, a new dwarf galaxy, Crater\,II, was discovered in the vicinity of Crater, aligned with the same great circle as the other three objects (\citealt{Torre2016}). This dwarf galaxy is very extended,  $R_{\rm e}=1.1\rm \,kpc$, considering its relatively faint magnitude of M$_{\rm v}=-8$. Its apparent angular separation from Crater is only $\sim 8^{\circ}$, which is equivalent to a distance of 20\,kpc at a distance of 145\,kpc. In addition, the derived metallicity of -1.7\,dex for the Crater\,II dwarf from \citet{Torre2016} is consistent with the values derived for Crater, giving further reason to speculate that Crater 2 is the dwarf galaxy that has potentially hosted Crater, and what we see now are the leftovers of a satellite group that forms a narrow tidal debris stream.

The hypothesis that Crater was stripped from a dwarf galaxy is further supported by the existence of extended GCs in other dwarf galaxies (e.g \citealt{Georgiev2009, Dacosta2009}), in addition to their presence in the outer Milky Way and M31 (e.g. \citealt{Huxor2005}). The extended cluster in the dwarf elliptical Sc22 in the Sculptor group has a half-light radius of 22\,pc making it one of the largest known GCs that is associated with a dwarf galaxy (\citealt{Dacosta2009}). In addition, its metallicity was found to be [Fe/H]=$-$1.7\, dex making it similarly metal-poor as Crater. In this work, it is suggested that GCs in dwarf galaxies can form in two modes, one "normal" mode and an extended mode with half-light radii of larger than 10\,pc. It is suggested that an environment with low stellar density is the requirement for the formation of clusters in the extended mode, which supports the view that Crater originated in another dwarf galaxy and has been stripped from it.

\subsection{Crater among Milky Way halo GCs and dwarf galaxies}
In the Milky Way halo, globular clusters that are at larger galactocentric distances than 100\,kpc are extremely rare. Only six Milky Way GCs can be found further out than 50\,kpc, and of those two lie at distances larger than 100\,kpc (\citealt{Harris1996}). 
The population of globular clusters in the outer Milky Way halo is on average younger and has a lower central stellar density than the inner halo GCs (\citealt{Dotter2010}). But among those "young" outer halo GCs, with an age of 7\,Gyr, Crater is still considerably younger than any other GC at galactocentric distances larger than 100\,kpc, which all have ages of 10\,Gyr or larger. 
Several authors have suggested that these young outer halo GCs were accreted onto the Milky Way along with their dwarf galaxy hosts, whereas the inner globular clusters formed via direct early collapse of the inner halo (\citealt{Cote2000, Lee2007, Forbes2010}). As already mentioned, it was suggested in \citet{Weisz2015} that Crater was accreted onto the Milky Way as a GC within a dwarf galaxy, from which it was later stripped, consistent with the proposed origin of the other outer halo GCs.

While Crater's M/L is elevated compared to models of a purely baryonic stellar system, there are cases in the literature where higher dynamical M/L ratios were derived for objects that are clearly GCs and not dwarf galaxies. One comparable object is the globular cluster NGC\,2257 in the Large Magellanic Cloud, which has a high dynamical mass-to-light ratio of 10.4 (\citealt{Laughlin2005}), similar to what we measure for Crater. This GC was also measured to be metal-poor with [Fe/H]=$-$1.59\, dex. This is comparable to what we measure for Crater. The reason for the elevated mass-to-light ratio of NGC\,2257 is still unknown. 

Another intriguing cluster that exhibits similar properties as Crater is Lindsay 38 (\citealt{Glatt2008}), one of the most remote clusters of the SMC with a distance of 68\,kpc to the SMC. With an age of 6.5\,Gyr and a metallicity of [Fe/H]=$-$1.59\, dex it is an intermediate age and metal-poor cluster. Its remote distance, size, age and metallicity are all similar to Crater's properties. 

The most similar objects to Crater are thus the intermediate-age extended clusters of the LMC/SMC. Therefore, the possibility that Crater was formed as a member of the LMC/SMC and later stripped from them appears also like a viable formation channel. 
Crater's position in the all-sky plot is close enough to the LMC-SMC orbit to be in principle consistent with a stripping scenario (see e.g. Fig. 1 \citealt{Pawlowski2015}). Nevertheless, the LMC-SMC orbit is oriented among the Vast Polar Structure of satellite galaxies of the Milky Way, and thus we expect many satellite objects to be aligned with this plane. Crater's true, deprojected position is $\sim$150\,kpc away from the LMC, which makes a stripping origin from the LMC relatively unlikely if we assume the LMC is only on its first perigalactic passage (e.g. \citealt{Besla2007, Kalli2013}).

In \citet{Jethwa2016} the distribution of the satellites of the Magellanic Clouds was modelled dynamically. In their likelihood distribution, Crater would be located just at the border of their projected satellite likelihood distribution. Thus Craters position is consistent with this model, although the probability is relatively low.\citet{Deason2015} use dark matter simulations to predict the fraction of Milky Way satellites at a certain distance that were once satellites bound to the LMC. According to their simulations,
the fraction of satellites that were bound to the LMC would amount to $\sim$10\% at a present-day distance of 150\,kpc. In summary, current models of the LMC-SMC satellite distributions cannot exclude that Crater originated from either LMC or SMC, but dynamical models predict a low probability for a LMC-SMC origin of Crater. However, until proper motion measurements will be conducted, no firm conclusions can be reached on its connection to the Magellanic System.

\section{Summary}
Our MUSE observations of Crater revealed that it has a systemic radial velocity of  $v_{\rm sys}=148.18^{\rm +1.08}_{\rm -1.15}\rm \,km\,s^{-1}$.
Its most likely velocity dispersion is $\sigma_{\rm v}=2.04^{\rm +2.19}_{\rm -1.06}  \rm \,km\,s^{-1}$, which leads to a total dynamical mass of the system, assuming dynamical equilibrium, of  $M_{\rm tot}=1.50^{+4.9}_{-1.2}\cdot 10^{\rm 5}M_{\odot}$. This dynamical mass implies a mass-to-light ratio of M/L$_{\rm V}$=8.52$^{+28.0}_{-6.5}\, M_{\odot}/L_{\odot}$, which is consistent with a purely baryonic stellar population within its errors and no significant evidence for the presence dark matter is found.

We conclude that our MUSE results, as well as the results of the other recent work on Crater (\citealt{Weisz2015, Kirby2015}), 
all strongly support Crater to be a globular cluster and not a dwarf galaxy.  Especially the
deep, high-quality CMD of \cite{Weisz2015} shows no evidence of more recent star formation. 
Our spectroscopy excludes membership of two of the luminous blue stars, and the data as well as the position 
of the third blue star make its membership quite unlikely. Even if this blue star were a genuine member it is difficult 
to explain why there would be only one clearly identifiable, luminous star left over from a more recent episode of star-formation.
The comparatively young age of Crater makes it a direct counterpart of the intermediate-age clusters in the SMC, 
many of which are located at large distances from the SMC's centre and could be stripped relatively easily from the SMC during an interaction.

The young age - with no obvious presence of an old population - is another argument against a dwarf galaxy.
 All dwarf galaxies that are sufficiently close to be resolved into stars have been found to contain old populations, without
 exception (\citealt{Grebel2004}). The ultrafaint dwarf spheroidals may have had their star formation
 truncated by reionization and never recovered their gas to undergo more recent star-formation. 
 Assuming that Crater formed as an isolated dwarf galaxy, the question remains, how such a low-mass object 
 would have managed to avoid SF at early times and to have been able to retain its gas to have only one single burst of star-formation 7\,Gyr ago, well after reionization?
In addition, its dynamical M/L is much lower than that of any other dSph of that luminosity (e.g. \citealt{Mcconn2012}).
Although these are not definitive exclusion arguments, they show that, if Crater had formed as a genuine dwarf galaxy, 
it would defy our understanding of dSphs and of dwarf galaxies in general. Thus we conclude that our findings strongly 
support that Crater is a faint intermediate-age outer halo globular cluster and not a dwarf galaxy.

\section*{Acknowledgements}
We thank our referee for helpful comments and suggestions that improve this paper, and made us aware of the first discovery of Crater by an amateur astronomer. We would like to thank Vasily Belokurov for providing us with his photometric dataset of Crater. We would also like to thank Sebastian Kamann for providing us with his Pampelmuse software and answering our questions about its implementation. We would like to thank Marcel Pawlowski for helpful discussions. Based on observations collected at the European Organisation for Astronomical Research in the Southern Hemisphere under ESO programme 094.D-0880. T.R. acknowledges support from the BASAL Centro de Astrof\'{\i}sica y Tecnologias Afines (CATA) PFB-06/2007 and from the ESO Scientific Visitor Programme. M.L.M.C acknowledges financial support from the European Research Council (ERC-StG-335936). M.J.F acknowledges support from the German Research Foundation (DFG) via Emmy Noether Grant Ko 4161/1. EKG acknowledges support by Sonderforschungsbereich SFB 881 "The Milky Way System" (subproject A2) of the German Research Foundation (DFG).

%%%%%%%%%%%%%%%%%%%%%%%%%%%%%%%%%%%%%%%%%%%%%%%%%%



\bibliographystyle{mnras}
\bibliography{bib_crater} % if your bibtex file is called example.bib

\begin{thebibliography}{}
\makeatletter
\relax
\def\mn@urlcharsother{\let\do\@makeother \do\$\do\&\do\#\do\^\do\_\do\%\do\~}
\def\mn@doi{\begingroup\mn@urlcharsother \@ifnextchar [ {\mn@doi@}
  {\mn@doi@[]}}
\def\mn@doi@[#1]#2{\def\@tempa{#1}\ifx\@tempa\@empty \href
  {http://dx.doi.org/#2} {doi:#2}\else \href {http://dx.doi.org/#2} {#1}\fi
  \endgroup}
\def\mn@eprint#1#2{\mn@eprint@#1:#2::\@nil}
\def\mn@eprint@arXiv#1{\href {http://arxiv.org/abs/#1} {{\tt arXiv:#1}}}
\def\mn@eprint@dblp#1{\href {http://dblp.uni-trier.de/rec/bibtex/#1.xml}
  {dblp:#1}}
\def\mn@eprint@#1:#2:#3:#4\@nil{\def\@tempa {#1}\def\@tempb {#2}\def\@tempc
  {#3}\ifx \@tempc \@empty \let \@tempc \@tempb \let \@tempb \@tempa \fi \ifx
  \@tempb \@empty \def\@tempb {arXiv}\fi \@ifundefined
  {mn@eprint@\@tempb}{\@tempb:\@tempc}{\expandafter \expandafter \csname
  mn@eprint@\@tempb\endcsname \expandafter{\@tempc}}}

\bibitem[\protect\citeauthoryear{{Bacon} et~al.,}{{Bacon}
  et~al.}{2010}]{Bacon2010}
{Bacon} R.,  et~al., 2010, in Society of Photo-Optical Instrumentation
  Engineers (SPIE) Conference Series. p.~8, \mn@doi{10.1117/12.856027}

\bibitem[\protect\citeauthoryear{{Baumgardt}, {Grebel}  \&
  {Kroupa}}{{Baumgardt} et~al.}{2005}]{Baumgardt2005}
{Baumgardt} H.,  {Grebel} E.~K.,   {Kroupa} P.,  2005, \mn@doi [\mnras]
  {10.1111/j.1745-3933.2005.00021.x}, \href
  {http://esoads.eso.org/abs/2005MNRAS.359L...1B} {359, L1}

\bibitem[\protect\citeauthoryear{{Baumgardt}, {Parmentier}, {Gieles}  \&
  {Vesperini}}{{Baumgardt} et~al.}{2010}]{Baumgardt2010}
{Baumgardt} H.,  {Parmentier} G.,  {Gieles} M.,   {Vesperini} E.,  2010,
  \mn@doi [\mnras] {10.1111/j.1365-2966.2009.15758.x}, \href
  {http://adsabs.harvard.edu/abs/2010MNRAS.401.1832B} {401, 1832}

\bibitem[\protect\citeauthoryear{{Belokurov} et~al.,}{{Belokurov}
  et~al.}{2007}]{Belokurov2007}
{Belokurov} V.,  et~al., 2007, \mn@doi [\apj] {10.1086/509718}, \href
  {http://adsabs.harvard.edu/abs/2007ApJ...654..897B} {654, 897}

\bibitem[\protect\citeauthoryear{{Belokurov} et~al.,}{{Belokurov}
  et~al.}{2008}]{Belokurov2008}
{Belokurov} V.,  et~al., 2008, \mn@doi [\apjl] {10.1086/592962}, \href
  {http://adsabs.harvard.edu/abs/2008ApJ...686L..83B} {686, L83}

\bibitem[\protect\citeauthoryear{{Belokurov}, {Irwin}, {Koposov}, {Evans},
  {Gonzalez-Solares}, {Metcalfe}  \& {Shanks}}{{Belokurov}
  et~al.}{2014}]{Belokurov2014}
{Belokurov} V.,  {Irwin} M.~J.,  {Koposov} S.~E.,  {Evans} N.~W.,
  {Gonzalez-Solares} E.,  {Metcalfe} N.,   {Shanks} T.,  2014, \mn@doi [\mnras]
  {10.1093/mnras/stu626}, \href
  {http://adsabs.harvard.edu/abs/2014MNRAS.441.2124B} {441, 2124}

\bibitem[\protect\citeauthoryear{{Besla}, {Kallivayalil}, {Hernquist},
  {Robertson}, {Cox}, {van der Marel}  \& {Alcock}}{{Besla}
  et~al.}{2007}]{Besla2007}
{Besla} G.,  {Kallivayalil} N.,  {Hernquist} L.,  {Robertson} B.,  {Cox} T.~J.,
   {van der Marel} R.~P.,   {Alcock} C.,  2007, \mn@doi [\apj]
  {10.1086/521385}, \href {http://adsabs.harvard.edu/abs/2007ApJ...668..949B}
  {668, 949}

\bibitem[\protect\citeauthoryear{{Bonifacio}, {Caffau}, {Zaggia}, {Fran{\c
  c}ois}, {Sbordone}, {Andrievsky}  \& {Korotin}}{{Bonifacio}
  et~al.}{2015}]{Bonifacio2015}
{Bonifacio} P.,  {Caffau} E.,  {Zaggia} S.,  {Fran{\c c}ois} P.,  {Sbordone}
  L.,  {Andrievsky} S.~M.,   {Korotin} S.~A.,  2015, \mn@doi [\aap]
  {10.1051/0004-6361/201526366}, \href
  {http://adsabs.harvard.edu/abs/2015A\%26A...579L...6B} {579, L6}

\bibitem[\protect\citeauthoryear{{Campbell} et~al.,}{{Campbell}
  et~al.}{2016}]{campbell2016}
{Campbell} D.~J.~R.,  et~al., 2016, preprint, \href
  {http://adsabs.harvard.edu/abs/2016arXiv160304443C} {} (\mn@eprint {arXiv}
  {1603.04443})

\bibitem[\protect\citeauthoryear{{Cenarro}, {Cardiel}, {Gorgas}, {Peletier},
  {Vazdekis}  \& {Prada}}{{Cenarro} et~al.}{2001}]{Cenarro2001}
{Cenarro} A.~J.,  {Cardiel} N.,  {Gorgas} J.,  {Peletier} R.~F.,  {Vazdekis}
  A.,   {Prada} F.,  2001, \mn@doi [\mnras] {10.1046/j.1365-8711.2001.04688.x},
  \href {http://ukads.nottingham.ac.uk/abs/2001MNRAS.326..959C} {326, 959}

\bibitem[\protect\citeauthoryear{{Collins} et~al.,}{{Collins}
  et~al.}{2013}]{Collins2013}
{Collins} M.~L.~M.,  et~al., 2013, \mn@doi [\apj]
  {10.1088/0004-637X/768/2/172}, \href
  {http://adsabs.harvard.edu/abs/2013ApJ...768..172C} {768, 172}

\bibitem[\protect\citeauthoryear{{C{\^o}t{\'e}}, {Marzke}, {West}  \&
  {Minniti}}{{C{\^o}t{\'e}} et~al.}{2000}]{Cote2000}
{C{\^o}t{\'e}} P.,  {Marzke} R.~O.,  {West} M.~J.,   {Minniti} D.,  2000,
  \mn@doi [\apj] {10.1086/308709}, \href
  {http://adsabs.harvard.edu/abs/2000ApJ...533..869C} {533, 869}

\bibitem[\protect\citeauthoryear{{Da Costa}, {Grebel}, {Jerjen}, {Rejkuba}  \&
  {Sharina}}{{Da Costa} et~al.}{2009}]{Dacosta2009}
{Da Costa} G.~S.,  {Grebel} E.~K.,  {Jerjen} H.,  {Rejkuba} M.,   {Sharina}
  M.~E.,  2009, \mn@doi [\aj] {10.1088/0004-6256/137/5/4361}, \href
  {http://adsabs.harvard.edu/abs/2009AJ....137.4361D} {137, 4361}

\bibitem[\protect\citeauthoryear{{Deason}, {Wetzel}, {Garrison-Kimmel}  \&
  {Belokurov}}{{Deason} et~al.}{2015}]{Deason2015}
{Deason} A.~J.,  {Wetzel} A.~R.,  {Garrison-Kimmel} S.,   {Belokurov} V.,
  2015, \mn@doi [\mnras] {10.1093/mnras/stv1939}, \href
  {http://adsabs.harvard.edu/abs/2015MNRAS.453.3568D} {453, 3568}

\bibitem[\protect\citeauthoryear{{Dotter} et~al.,}{{Dotter}
  et~al.}{2010}]{Dotter2010}
{Dotter} A.,  et~al., 2010, \mn@doi [\apj] {10.1088/0004-637X/708/1/698}, \href
  {http://adsabs.harvard.edu/abs/2010ApJ...708..698D} {708, 698}

\bibitem[\protect\citeauthoryear{{Ferraro} et~al.,}{{Ferraro}
  et~al.}{2012}]{Ferraro2012}
{Ferraro} F.~R.,  et~al., 2012, \mn@doi [\nat] {10.1038/nature11686}, \href
  {http://adsabs.harvard.edu/abs/2012Natur.492..393F} {492, 393}

\bibitem[\protect\citeauthoryear{{Forbes} \& {Bridges}}{{Forbes} \&
  {Bridges}}{2010}]{Forbes2010}
{Forbes} D.~A.,  {Bridges} T.,  2010, \mn@doi [\mnras]
  {10.1111/j.1365-2966.2010.16373.x}, \href
  {http://adsabs.harvard.edu/abs/2010MNRAS.404.1203F} {404, 1203}

\bibitem[\protect\citeauthoryear{{Frank}, {Hilker}, {Baumgardt},
  {C{\^o}t{\'e}}, {Grebel}, {Haghi}, {K{\"u}pper}  \& {Djorgovski}}{{Frank}
  et~al.}{2012}]{Frank2012}
{Frank} M.~J.,  {Hilker} M.,  {Baumgardt} H.,  {C{\^o}t{\'e}} P.,  {Grebel}
  E.~K.,  {Haghi} H.,  {K{\"u}pper} A.~H.~W.,   {Djorgovski} S.~G.,  2012,
  \mn@doi [\mnras] {10.1111/j.1365-2966.2012.21105.x}, \href
  {http://adsabs.harvard.edu/abs/2012MNRAS.423.2917F} {423, 2917}

\bibitem[\protect\citeauthoryear{{Georgiev}, {Puzia}, {Hilker}  \&
  {Goudfrooij}}{{Georgiev} et~al.}{2009}]{Georgiev2009}
{Georgiev} I.~Y.,  {Puzia} T.~H.,  {Hilker} M.,   {Goudfrooij} P.,  2009,
  \mn@doi [\mnras] {10.1111/j.1365-2966.2008.14104.x}, \href
  {http://adsabs.harvard.edu/abs/2009MNRAS.392..879G} {392, 879}

\bibitem[\protect\citeauthoryear{{Gilmore}, {Wilkinson}, {Wyse}, {Kleyna},
  {Koch}, {Evans}  \& {Grebel}}{{Gilmore} et~al.}{2007}]{Gilmore2007}
{Gilmore} G.,  {Wilkinson} M.~I.,  {Wyse} R.~F.~G.,  {Kleyna} J.~T.,  {Koch}
  A.,  {Evans} N.~W.,   {Grebel} E.~K.,  2007, \mn@doi [\apj] {10.1086/518025},
  \href {http://adsabs.harvard.edu/abs/2007ApJ...663..948G} {663, 948}

\bibitem[\protect\citeauthoryear{{Glatt} et~al.,}{{Glatt}
  et~al.}{2008}]{Glatt2008}
{Glatt} K.,  et~al., 2008, \mn@doi [\aj] {10.1088/0004-6256/136/4/1703}, \href
  {http://adsabs.harvard.edu/abs/2008AJ....136.1703G} {136, 1703}

\bibitem[\protect\citeauthoryear{{Glatt} et~al.,}{{Glatt}
  et~al.}{2009}]{Glatt2009}
{Glatt} K.,  et~al., 2009, \mn@doi [\aj] {10.1088/0004-6256/138/5/1403}, \href
  {http://adsabs.harvard.edu/abs/2009AJ....138.1403G} {138, 1403}

\bibitem[\protect\citeauthoryear{{Grebel} \& {Gallagher}}{{Grebel} \&
  {Gallagher}}{2004}]{Grebel2004}
{Grebel} E.~K.,  {Gallagher} III J.~S.,  2004, \mn@doi [\apjl]
  {10.1086/423339}, \href {http://adsabs.harvard.edu/abs/2004ApJ...610L..89G}
  {610, L89}

\bibitem[\protect\citeauthoryear{{Hanuschik}}{{Hanuschik}}{2003}]{Hanuschik2003}
{Hanuschik} R.~W.,  2003, \mn@doi [\aap] {10.1051/0004-6361:20030885}, \href
  {http://adsabs.harvard.edu/abs/2003A26A...407.1157H} {407, 1157}

\bibitem[\protect\citeauthoryear{{Harris}}{{Harris}}{1996}]{Harris1996}
{Harris} W.~E.,  1996, \mn@doi [\aj] {10.1086/118116}, \href
  {http://adsabs.harvard.edu/abs/1996AJ....112.1487H} {112, 1487}

\bibitem[\protect\citeauthoryear{{Huxor}, {Tanvir}, {Irwin}, {Ibata},
  {Collett}, {Ferguson}, {Bridges}  \& {Lewis}}{{Huxor}
  et~al.}{2005}]{Huxor2005}
{Huxor} A.~P.,  {Tanvir} N.~R.,  {Irwin} M.~J.,  {Ibata} R.,  {Collett} J.~L.,
  {Ferguson} A.~M.~N.,  {Bridges} T.,   {Lewis} G.~F.,  2005, \mn@doi [\mnras]
  {10.1111/j.1365-2966.2005.09086.x}, \href
  {http://adsabs.harvard.edu/abs/2005MNRAS.360.1007H} {360, 1007}

\bibitem[\protect\citeauthoryear{{Ibata}, {Sollima}, {Nipoti}, {Bellazzini},
  {Chapman}  \& {Dalessandro}}{{Ibata} et~al.}{2011}]{Ibata2011}
{Ibata} R.,  {Sollima} A.,  {Nipoti} C.,  {Bellazzini} M.,  {Chapman} S.~C.,
  {Dalessandro} E.,  2011, \mn@doi [\apj] {10.1088/0004-637X/738/2/186}, \href
  {http://esoads.eso.org/abs/2011ApJ...738..186I} {738, 186}

\bibitem[\protect\citeauthoryear{{Jethwa}, {Erkal}  \& {Belokurov}}{{Jethwa}
  et~al.}{2016}]{Jethwa2016}
{Jethwa} P.,  {Erkal} D.,   {Belokurov} V.,  2016, preprint, \href
  {http://adsabs.harvard.edu/abs/2016arXiv160304420J} {} (\mn@eprint {arXiv}
  {1603.04420})

\bibitem[\protect\citeauthoryear{{Jordi} et~al.,}{{Jordi}
  et~al.}{2009}]{Jordi2009}
{Jordi} K.,  et~al., 2009, \mn@doi [\aj] {10.1088/0004-6256/137/6/4586}, \href
  {http://adsabs.harvard.edu/abs/2009AJ....137.4586J} {137, 4586}

\bibitem[\protect\citeauthoryear{{Kallivayalil}, {van der Marel}, {Besla},
  {Anderson}  \& {Alcock}}{{Kallivayalil} et~al.}{2013}]{Kalli2013}
{Kallivayalil} N.,  {van der Marel} R.~P.,  {Besla} G.,  {Anderson} J.,
  {Alcock} C.,  2013, \mn@doi [\apj] {10.1088/0004-637X/764/2/161}, \href
  {http://adsabs.harvard.edu/abs/2013ApJ...764..161K} {764, 161}

\bibitem[\protect\citeauthoryear{{Kamann}, {Wisotzki}  \& {Roth}}{{Kamann}
  et~al.}{2013}]{Kamann2013}
{Kamann} S.,  {Wisotzki} L.,   {Roth} M.~M.,  2013, \mn@doi [\aap]
  {10.1051/0004-6361/201220476}, \href
  {http://adsabs.harvard.edu/abs/2013A26A...549A..71K} {549, A71}

\bibitem[\protect\citeauthoryear{{Kirby}, {Simon}  \& {Cohen}}{{Kirby}
  et~al.}{2015}]{Kirby2015}
{Kirby} E.~N.,  {Simon} J.~D.,   {Cohen} J.~G.,  2015, \mn@doi [\apj]
  {10.1088/0004-637X/810/1/56}, \href
  {http://adsabs.harvard.edu/abs/2015ApJ...810...56K} {810, 56}

\bibitem[\protect\citeauthoryear{{K{\"u}pper}, {Mieske}  \&
  {Kroupa}}{{K{\"u}pper} et~al.}{2011}]{Kuepper2011}
{K{\"u}pper} A.~H.~W.,  {Mieske} S.,   {Kroupa} P.,  2011, \mn@doi [\mnras]
  {10.1111/j.1365-2966.2010.18178.x}, \href
  {http://adsabs.harvard.edu/abs/2011MNRAS.413..863K} {413, 863}

\bibitem[\protect\citeauthoryear{{Laevens} et~al.,}{{Laevens}
  et~al.}{2014}]{Laevens2014}
{Laevens} B.~P.~M.,  et~al., 2014, \mn@doi [\apjl]
  {10.1088/2041-8205/786/1/L3}, \href
  {http://adsabs.harvard.edu/abs/2014ApJ...786L...3L} {786, L3}

\bibitem[\protect\citeauthoryear{{Laevens} et~al.,}{{Laevens}
  et~al.}{2015}]{Laevens2015}
{Laevens} B.~P.~M.,  et~al., 2015, \mn@doi [\apj] {10.1088/0004-637X/813/1/44},
  \href {http://adsabs.harvard.edu/abs/2015ApJ...813...44L} {813, 44}

\bibitem[\protect\citeauthoryear{{Lee}, {Gim}  \& {Casetti-Dinescu}}{{Lee}
  et~al.}{2007}]{Lee2007}
{Lee} Y.-W.,  {Gim} H.~B.,   {Casetti-Dinescu} D.~I.,  2007, \mn@doi [\apjl]
  {10.1086/518653}, \href {http://adsabs.harvard.edu/abs/2007ApJ...661L..49L}
  {661, L49}

\bibitem[\protect\citeauthoryear{{Maraston}}{{Maraston}}{2005}]{Maraston2005}
{Maraston} C.,  2005, \mn@doi [\mnras] {10.1111/j.1365-2966.2005.09270.x},
  \href {http://adsabs.harvard.edu/abs/2005MNRAS.362..799M} {362, 799}

\bibitem[\protect\citeauthoryear{{Martin}, {Ibata}, {Chapman}, {Irwin}  \&
  {Lewis}}{{Martin} et~al.}{2007}]{Martin2007}
{Martin} N.~F.,  {Ibata} R.~A.,  {Chapman} S.~C.,  {Irwin} M.,   {Lewis} G.~F.,
   2007, \mn@doi [\mnras] {10.1111/j.1365-2966.2007.12055.x}, \href
  {http://adsabs.harvard.edu/abs/2007MNRAS.380..281M} {380, 281}

\bibitem[\protect\citeauthoryear{{Martin} et~al.,}{{Martin}
  et~al.}{2015}]{Martin2015c}
{Martin} N.~F.,  et~al., 2015, \mn@doi [\apjl] {10.1088/2041-8205/804/1/L5},
  \href {http://adsabs.harvard.edu/abs/2015ApJ...804L...5M} {804, L5}

\bibitem[\protect\citeauthoryear{{Martin} et~al.,}{{Martin}
  et~al.}{2016a}]{Martin2015b}
{Martin} N.~F.,  et~al., 2016a, \mn@doi [\mnras] {10.1093/mnrasl/slw013}, \href
  {http://adsabs.harvard.edu/abs/2016MNRAS.458L..59M} {458, L59}

\bibitem[\protect\citeauthoryear{{Martin} et~al.,}{{Martin}
  et~al.}{2016b}]{Martin2015a}
{Martin} N.~F.,  et~al., 2016b, \mn@doi [\apj] {10.3847/0004-637X/818/1/40},
  \href {http://adsabs.harvard.edu/abs/2016ApJ...818...40M} {818, 40}

\bibitem[\protect\citeauthoryear{{McConnachie}}{{McConnachie}}{2012}]{Mcconn2012}
{McConnachie} A.~W.,  2012, \mn@doi [\aj] {10.1088/0004-6256/144/1/4}, \href
  {http://adsabs.harvard.edu/abs/2012AJ....144....4M} {144, 4}

\bibitem[\protect\citeauthoryear{{McConnachie} \& {C{\^o}t{\'e}}}{{McConnachie}
  \& {C{\^o}t{\'e}}}{2010}]{Mcconn2010}
{McConnachie} A.~W.,  {C{\^o}t{\'e}} P.,  2010, \mn@doi [\apjl]
  {10.1088/2041-8205/722/2/L209}, \href
  {http://adsabs.harvard.edu/abs/2010ApJ...722L.209M} {722, L209}

\bibitem[\protect\citeauthoryear{{McLaughlin} \& {van der Marel}}{{McLaughlin}
  \& {van der Marel}}{2005}]{Laughlin2005}
{McLaughlin} D.~E.,  {van der Marel} R.~P.,  2005, \mn@doi [\apjs]
  {10.1086/497429}, \href {http://cdsads.u-strasbg.fr/abs/2005ApJS..161..304M}
  {161, 304}

\bibitem[\protect\citeauthoryear{{Milgrom}}{{Milgrom}}{1983}]{Milgrom1983}
{Milgrom} M.,  1983, \mn@doi [\apj] {10.1086/161130}, \href
  {http://adsabs.harvard.edu/abs/1983ApJ...270..365M} {270, 365}

\bibitem[\protect\citeauthoryear{{Milone} et~al.,}{{Milone}
  et~al.}{2012}]{Milone2012}
{Milone} A.~P.,  et~al., 2012, \mn@doi [\aap] {10.1051/0004-6361/201016384},
  \href {http://adsabs.harvard.edu/abs/2012A%26A...540A..16M} {540, A16}

\bibitem[\protect\citeauthoryear{{Misgeld} \& {Hilker}}{{Misgeld} \&
  {Hilker}}{2011}]{Misgeld2011}
{Misgeld} I.,  {Hilker} M.,  2011, \mn@doi [\mnras]
  {10.1111/j.1365-2966.2011.18669.x}, \href
  {http://adsabs.harvard.edu/abs/2011MNRAS.414.3699M} {414, 3699}

\bibitem[\protect\citeauthoryear{{Padoan}, {Jimenez}  \& {Jones}}{{Padoan}
  et~al.}{1997}]{Padoan1997}
{Padoan} P.,  {Jimenez} R.,   {Jones} B.,  1997, \mn@doi [\mnras]
  {10.1093/mnras/285.4.711}, \href
  {http://adsabs.harvard.edu/abs/1997MNRAS.285..711P} {285, 711}

\bibitem[\protect\citeauthoryear{{Paust} et~al.,}{{Paust}
  et~al.}{2010}]{Paust2010}
{Paust} N.~E.~Q.,  et~al., 2010, \mn@doi [\aj] {10.1088/0004-6256/139/2/476},
  \href {http://adsabs.harvard.edu/abs/2010AJ....139..476P} {139, 476}

\bibitem[\protect\citeauthoryear{{Pawlowski}, {McGaugh}  \&
  {Jerjen}}{{Pawlowski} et~al.}{2015}]{Pawlowski2015}
{Pawlowski} M.~S.,  {McGaugh} S.~S.,   {Jerjen} H.,  2015, \mn@doi [\mnras]
  {10.1093/mnras/stv1588}, \href
  {http://adsabs.harvard.edu/abs/2015MNRAS.453.1047P} {453, 1047}

\bibitem[\protect\citeauthoryear{{Peebles}}{{Peebles}}{1984}]{Peebles1984}
{Peebles} P.~J.~E.,  1984, \mn@doi [\apj] {10.1086/161714}, \href
  {http://adsabs.harvard.edu/abs/1984ApJ...277..470P} {277, 470}

\bibitem[\protect\citeauthoryear{{Robin}, {Reyl{\'e}}, {Derri{\`e}re}  \&
  {Picaud}}{{Robin} et~al.}{2003}]{Besancon2003}
{Robin} A.~C.,  {Reyl{\'e}} C.,  {Derri{\`e}re} S.,   {Picaud} S.,  2003,
  \mn@doi [\aap] {10.1051/0004-6361:20031117}, \href
  {http://adsabs.harvard.edu/abs/2003A26A...409..523R} {409, 523}

\bibitem[\protect\citeauthoryear{{Sollima}, {Bellazzini}  \& {Lee}}{{Sollima}
  et~al.}{2012}]{Sollima2012}
{Sollima} A.,  {Bellazzini} M.,   {Lee} J.-W.,  2012, \mn@doi [\apj]
  {10.1088/0004-637X/755/2/156}, \href
  {http://adsabs.harvard.edu/abs/2012ApJ...755..156S} {755, 156}

\bibitem[\protect\citeauthoryear{{Torrealba}, {Koposov}, {Belokurov}  \&
  {Irwin}}{{Torrealba} et~al.}{2016}]{Torre2016}
{Torrealba} G.,  {Koposov} S.~E.,  {Belokurov} V.,   {Irwin} M.,  2016,
  preprint, \href {http://adsabs.harvard.edu/abs/2016arXiv160107178T} {}
  (\mn@eprint {arXiv} {1601.07178})

\bibitem[\protect\citeauthoryear{{Weilbacher}, {Streicher}, {Urrutia}, {Jarno},
  {P{\'e}contal-Rousset}, {Bacon}  \& {B{\"o}hm}}{{Weilbacher}
  et~al.}{2012}]{Weilbacher2012}
{Weilbacher} P.~M.,  {Streicher} O.,  {Urrutia} T.,  {Jarno} A.,
  {P{\'e}contal-Rousset} A.,  {Bacon} R.,   {B{\"o}hm} P.,  2012, in Software
  and Cyberinfrastructure for Astronomy II. p. 84510B,
  \mn@doi{10.1117/12.925114}

\bibitem[\protect\citeauthoryear{{Weisz} et~al.,}{{Weisz}
  et~al.}{2015}]{Weisz2015}
{Weisz} D.~R.,  et~al., 2015, preprint, \href
  {http://adsabs.harvard.edu/abs/2015arXiv151008533W} {} (\mn@eprint {arXiv}
  {1510.08533})

\bibitem[\protect\citeauthoryear{{Willman} \& {Strader}}{{Willman} \&
  {Strader}}{2012}]{Willman2012}
{Willman} B.,  {Strader} J.,  2012, \mn@doi [\aj] {10.1088/0004-6256/144/3/76},
  \href {http://adsabs.harvard.edu/abs/2012AJ....144...76W} {144, 76}

\bibitem[\protect\citeauthoryear{{Willman} et~al.,}{{Willman}
  et~al.}{2005}]{Willman2005}
{Willman} B.,  et~al., 2005, \mn@doi [\aj] {10.1086/430214}, \href
  {http://adsabs.harvard.edu/abs/2005AJ....129.2692W} {129, 2692}

\bibitem[\protect\citeauthoryear{{Wolf}, {Martinez}, {Bullock}, {Kaplinghat},
  {Geha}, {Mu{\~n}oz}, {Simon}  \& {Avedo}}{{Wolf} et~al.}{2010}]{Wolf2010}
{Wolf} J.,  {Martinez} G.~D.,  {Bullock} J.~S.,  {Kaplinghat} M.,  {Geha} M.,
  {Mu{\~n}oz} R.~R.,  {Simon} J.~D.,   {Avedo} F.~F.,  2010, \mn@doi [\mnras]
  {10.1111/j.1365-2966.2010.16753.x}, \href
  {http://adsabs.harvard.edu/abs/2010MNRAS.406.1220W} {406, 1220}

\bibitem[\protect\citeauthoryear{{Zucker} et~al.,}{{Zucker}
  et~al.}{2006a}]{Zucker2006}
{Zucker} D.~B.,  et~al., 2006a, \mn@doi [\apjl] {10.1086/505216}, \href
  {http://adsabs.harvard.edu/abs/2006ApJ...643L.103Z} {643, L103}

\bibitem[\protect\citeauthoryear{{Zucker} et~al.,}{{Zucker}
  et~al.}{2006b}]{Zucker2006b}
{Zucker} D.~B.,  et~al., 2006b, \mn@doi [\apjl] {10.1086/508628}, \href
  {http://adsabs.harvard.edu/abs/2006ApJ...650L..41Z} {650, L41}

\makeatother
\end{thebibliography}

%\begin{thebibliography}{99}
%\end{thebibliography}

%%%%%%%%%%%%%%%%%%%%%%%%%%%%%%%%%%%%%%%%%%%%%%%%%%

%%%%%%%%%%%%%%%%% APPENDICES %%%%%%%%%%%%%%%%%%%%%

\appendix

\section{Table of analysed stars}

\begin{table*} 

\caption{Radial velocities of extracted stars}{} % title of Table 

\vspace{5 mm}

\centering      % used for centering table 
\begin{tabular}{c c c c c c c c c c}  % centered columns (8 columns) 
\hline\hline    
Index  & R.A. (J2000) & Dec.(J2000) & $I_{\rm mag}$ & $(g-i)$ & radial velocity & membership probability \\
& & & (mag) & (mag) & (km s$^{-1}$) & \\
\hline 
      2 &        174.06963 &       -10.879207 &        17.34 &        1.70&        147.37 $\pm$        2.63   &     0.86 \\
       3 &        174.06443 &       -10.869809 &        17.44 &        2.63 &       -26.43 $\pm$        2.33 &  0.00 \\
       4 &        174.05541 &       -10.863862 &        17.82 &        2.44 &       0.58 $\pm$        2.41  & 0.00 \\
       5 &        174.05849 &       -10.873851 &        18.56 &        1.26 &        149.94 $\pm$        2.43 &   0.50 \\
       6 &        174.07254 &       -10.879141 &        19.08 &       0.31 &        222.26 $\pm$        3.67 &  0.00 \\
       7 &        174.08216 &       -10.876634 &        19.13 &       0.40 &        154.17 $\pm$        2.83   &   0.20 \\
       8 &        174.06698 &       -10.878484 &        18.94 &        1.11&        145.64 $\pm$        2.40   &  0.95 \\
       9 &        174.06640 &       -10.889202 &        18.98 &        2.15 &        110.56 $\pm$        4.30 & 2.65e-05 \\
      10 &        174.07103 &       -10.876154 &        19.60 &       0.97 &        145.76 $\pm$        2.68    &    0.78 \\
      11 &        174.06869 &       -10.878836 &        19.86 &       0.92&        144.74 $\pm$        3.92    &   0.91 \\
      12 &        174.07132 &       -10.886548 &        19.70 &        2.14 &       -73.73 $\pm$        48.77  &  1.9e-4 \\
      13 &        174.06278 &       -10.872687 &        20.24 &       0.63 &        151.56 $\pm$        4.83    &    0.65 \\
      14 &        174.06996 &       -10.871523 &        20.41 &       0.85 &        154.11 $\pm$        5.14  &    0.59 \\
      15 &        174.06862 &       -10.869076 &        20.46 &       0.87 &        151.86 $\pm$        3.92   &     0.48 \\
      17 &        174.07026 &       -10.876373 &        20.71 &       0.76 &        151.75 $\pm$        4.62   &     0.81 \\
      18 &        174.06220 &       -10.870687 &        20.79 &       0.67&        136.40 $\pm$        5.43    &    0.51 \\
      19 &        174.07804 &       -10.880047 &        20.77&       0.82 &        146.56 $\pm$        8.14    &    0.36 \\
      20 &        174.07586 &       -10.872134 &        20.78 &       0.82 &        171.85 $\pm$        7.37   &    0.11 \\
      21 &        174.07129 &       -10.871307 &        20.79 &       0.82 &        149.99 $\pm$        6.27   &     0.55 \\
      22 &        174.06518 &       -10.876647 &        20.91 &       0.57 &        162.10 $\pm$        15.91  &      0.89 \\
      23 &        174.06143 &       -10.871692 &        20.95 &       0.55 &        169.97 $\pm$        12.53  &      0.50 \\
      24 &        174.06155 &       -10.869075 &        20.96 &       0.59 &        130.26 $\pm$        11.89  &     0.41 \\
      25 &        174.08091 &       -10.880816 &        20.95 &       0.62 &        142.80 $\pm$        8.40   &     0.24 \\
      26 &        174.06890 &       -10.872272 &        21.01 &       0.64&        160.83 $\pm$        7.26    &    0.62 \\
      27 &        174.06919 &       -10.877074 &        21.01 &       0.65 &        146.75 $\pm$        4.66   &     0.89 \\
      28 &        174.05927 &       -10.866367 &        21.00 &       0.64 &        133.17 $\pm$        14.59   &     0.26 \\
      29 &        174.06988 &       -10.876175 &        21.07 &       0.58&        133.64 $\pm$        6.42  &     0.79 \\
      30 &        174.06878 &       -10.874268 &        21.11 &       0.58 &        145.61 $\pm$        9.94   &    0.81 \\
      33 &        174.06238 &       -10.873799 &        21.23 &       0.72 &        159.49 $\pm$        8.57    &    0.67 \\
      34 &        174.06709 &       -10.877142 &        21.30 &       0.72 &        135.83 $\pm$        6.17   &     0.91  \\
      35 &        174.07098 &       -10.882941 &        21.32 &       0.76 &        131.20 $\pm$        7.12   &   0.60 \\
      36 &        174.07060 &       -10.881660 &        21.32 &       0.84 &        141.87 $\pm$        10.09   &     0.73 \\
      37 &        174.06429 &       -10.875020 &        21.46 &       0.65 &        199.18$\pm$        28.65    &    0.75 \\
      38 &        174.06759 &       -10.876072 &        21.57 &       0.63&        136.73 $\pm$        7.42    &    0.89 \\
      39 &        174.06818 &       -10.876399 &        21.89 &       0.57 &        159.09 $\pm$        6.91  &      0.84 \\
      40 &        174.06847 &       -10.876633 &        21.90 &       0.61 &        143.17 $\pm$        6.72   &     0.91 \\
      47 &        174.07033 &       -10.870255 &        21.50 &        1.30 &       -131.02$\pm$        10.62  & 0.00 \\
      48 &        174.06635 &       -10.877323 &        22.20 &       0.23 &        146.44 $\pm$        22.87  &     0.95 \\
      49 &        174.06653 &       -10.876920 &        22.06 &       0.51 &        142.49$\pm$        9.17    &    0.94\\
      50 &        174.06811 &       -10.873349 &        22.09 &       0.57 &        176.90 $\pm$        20.49    &   0.71 \\
      51 &        174.07868 &       -10.877564 &        22.00 &       0.81 &        178.90 $\pm$        11.70    &   0.15\\
  %    99 &              NaN &              NaN &              NaN &              NaN &        163.30$\pm$        13.72 &	NaN 	 \\
 \hline
\\

\end{tabular} 

\label{tab:dat}  % is used to refer this table in the text 

\begin{flushleft} This table lists the identifier, the R.A. and DEC, the $I$ magnitude any $g-i$ colour form \citealt{Belokurov2014} and the radial velocity and its uncertainty. The last column gives the membership probability for each star calculated
according to eq. \ref{eq:prb} \end{flushleft}

\end{table*} 

%%%%%%%%%%%%%%%%%%%%%%%%%%%%%%%%%%%%%%%%%%%%%%%%%%

% Don't change these lines
\bsp	% typesetting comment
\label{lastpage}
\end{document}